\documentclass[12pt,nofootinbib,aps,amsmath,amssymb,preprint,showpacs]{revtex4}
\usepackage[T1]{fontenc} \usepackage[latin1]{inputenc}
\usepackage{amsmath,color,ulem} \usepackage{graphicx}
\usepackage{epsfig} \usepackage{amssymb}
\begin{document}

\title{Semiflexible Polymer Dynamics with a Bead-Spring Model}

\author{Gerard T. Barkema} \affiliation{Institute for Theoretical
Physics, Universiteit Utrecht, Leuvenlaan 4, 3584 CE Utrecht, The
Netherlands} \affiliation{Instituut-Lorentz, Universiteit Leiden,
Niels Bohrweg 2, 2333 CA Leiden, The Netherlands} \author{Debabrata
Panja} \affiliation{Institute for Theoretical Physics, Universiteit
Utrecht, Leuvenlaan 4, 3584 CE Utrecht, The Netherlands}
\author{J. M. J. van Leeuwen} \affiliation{Instituut-Lorentz,
Universiteit Leiden, Niels Bohrweg 2, 2333 CA Leiden, The Netherlands}

\begin{abstract} We study the dynamical properties of semiflexible
  polymers with a recently introduced bead-spring model. We focus on
  double-stranded DNA (dsDNA). The two parameters of the model, $T^*$
  and $\nu$, are chosen to match its experimental force-extension
  curve. In comparison to its groundstate value, the bead-spring
  Hamiltonian is approximated in the first order by the Hessian that
  is quadratic in the bead positions. The eigenmodes of the Hessian
  provide the longitudinal (stretching) and transverse (bending)
  eigenmodes of the polymer, and the corresponding eigenvalues match
  well with the established phenomenology of semiflexible polymers. At
  the Hessian approximation of the Hamiltonian, the polymer dynamics
  is linear. Using the longitudinal and transverse eigenmodes, for the
  linearized problem, we obtain analytical expressions of (i) the
  autocorrelation function of the end-to-end vector, (ii) the
  autocorrelation function of a bond (i.e., a spring, or a tangent)
  vector at the middle of the chain, and (iii) the mean-square
  displacement of a tagged bead in the middle of the chain, as sum
  over the contributions from the modes --- the so-called ``mode
  sums''. We also perform simulations with the full dynamics of the
  model. The simulations yield numerical values of the correlations
  functions (i-iii) that agree very well with the analytical
  expressions for the linearized dynamics. This does not however mean
  that the nonlinearities are not present. In fact, we also study the
  mean-square displacement of the longitudinal component of the
  end-to-end vector that showcases strong nonlinear effects in the
  polymer dynamics, and we identify at least an effective $t^{7/8}$
  power-law regime in its time-dependence. Nevertheless, in comparison
  to the full mean-square displacement of the end-to-end vector the
  nonlinear effects remain small at all times --- it is in this sense
  we state that our results demonstrate that the linearized dynamics
  suffices for dsDNA fragments that are shorter than or comparable to
  the persistence length. Our results are consistent with those of
  the wormlike chain (WLC) model, the commonly used descriptive tool
  of semiflexible polymers.
\end{abstract}

\maketitle

\section{Introduction\label{sec1}}

The last decades have witnessed a surge in research activities in the
physical properties of biopolymers, such as double-stranded DNA
(dsDNA), filamental actin (F-actin) and microtubules. Semiflexibility
is a common feature they share; as they preserve mechanical rigidity
over a range, characterized by the persistence length $l_p$, along
their contour. (E.g., for a dsDNA, F-actin and microtubules,
$l_p\sim50$ nm \cite{dnapersist1,dnapersist2,wang}, $\sim16 \mu$m
\cite{Factinpersist} and $\sim5$ mm \cite{micropersist} respectively.)
Mechanical properties of semiflexible polymers are well-captured by
the Kratky-Porod wormlike chain (WLC) model \cite{kratky}, wherein the
chain conformation is described by an inextensible and differentiable
curve. In general, the stretching modulus for semiflexible polymers is
far greater than their bending modulus, i.e., the chains are
effectively inextensible at the length for which the persistence
length is relevant.

The WLC model \cite{kratky}, its subsequent modifications
\cite{mods1,mods2,mods3,mods4}, and recent analyses
\cite{recent1,recent2, recent3, recent4, recent5, recent6} have been
very successful in describing static/mechanical properties of dsDNA,
such as its force-extension curve and the radial distribution function
of its end-to-end distance. For the study of dynamics, the WLC model
needs to be extended. This is done by a Langevin-like description of
sideways excursions of the WLC. The contour length of the polymer is
constrained in the original WLC model, and Lagrangian multipliers of a
varying degree of sophistication need to be implemented in order to
enforce a contour length that is either strictly fixed
\cite{liverpool1,liverpool2,liverpool3,liverpool4,kroy}, or fixed on
average \cite{winkler1,winkler2}.  This enforcement allows for the
calculation of many dynamical quantities theoretically. It also makes
it difficult to deal with motion along the WLC's contour \cite{kas},
which become relevant, e.g., in crowded conditions. In this context,
we note that the standard interpretation of the ``stored lengths'' for
semiflexible polymers are their (transverse) thermal undulations
\cite{seifert} (which are penalized). Although this interpretation is
consistent with the dynamics of stored lengths
\cite{seifert,hall1,hall2,hall3,hall4}, and allows one to intuitively
conceptualize chain tension along the contour, the interplay between
local extensibility and dynamics remains somewhat problematic, unlike
in bead-spring models.

In order to circumvent some of the difficulties, in a recent paper
(hereafter referred to as paper I) \cite{hans} two of us recently
introduced a bead-spring model for extensible semiflexible polymers in
Hamiltonian formulation, in the absence of hydrodynamic interactions.
In this paper we use this bead-spring model to study the dynamics of
polymer chains which are shorter than or of the order of the
persistence length. In particular, we have in mind the simulation of
the dynamics of crosslinked networks of semiflexible polymers such as
actin\cite{huisman}; first making the network configurations
continuous, and then discretizing them again, seems a bit indirect; we
prefer to work with a discrete model right from the start.

Paper I focused on dsDNA and matched the model to its force extension
curve in terms of two parameters, $T^*$ and $\nu$. In comparison to
its groundstate value, the bead-spring Hamiltonian is approximated in
the first order by the Hessian that is quadratic in the bead
positions. The eigenmodes of the Hessian provide the longitudinal
(stretching) and transverse (bending) eigenmodes of the polymer, and
the corresponding eigenvalues for the stretching (longitudinal) and
the bending (transverse) modes, indexed by $p$ ($p=1,2,\ldots,N$),
were shown to scale as $\zeta^l_p\sim(p/N)^2$ and $\zeta^t_p\sim
p^2(p-1)^2/N^4$ respectively for small $p$. These are the
characteristic signatures of semiflexible polymers \cite{farge}. In
this paper, also focused on dsDNA, we exploit the knowledge on these
modes to theoretically obtain some interesting dynamical quantities,
such as (i) the end-to-end vector autocorrelation function, (ii) the
correlation function of a bond vector in the middle of the chain, and
(iii) the mean-squared displacement (MSD) of a tagged bead.  In the
linear regime, where the modes can be taken as independent, our
(analytical) mode sum results for the above quantities (i-iii) are the
following.  The end-to-end vector autocorrelation function for the
chain decays in time as a stretched exponential with an exponent
$3/4$, crossing over to pure exponential decay at the terminal time
$\tau^*=(\zeta_2^t)^{-1}$; this exponent is partially shared with that
of the WLC \cite{liverpool4}. The autocorrelation of the orientation
of the middlemost bond vector decays in time in a similar manner, but
with an exponent $1/4$; this property, too, is partially shared with
the WLC \cite{liverpool4}. The mean-square displacement (MSD) of the
middle bead shows anomalous diffusion with an exponent $3/4$ until
time $\tau^*$, beyond which its motion becomes diffusive; this
property has been reported for the WLC \cite{kroy,farge}, as well from
experiments \cite{expt1,expt2,expt3}.

The full dynamics of semiflexible polymers in our model is obviously
nonlinear, and we also perform simulations of the full dynamics for
chains. We find that for dsDNA the mode sums agree remarkably well
with the numerical values obtained from simulations. This does not
however mean that the nonlinearities are not present. Indeed, we find
that the MSD of the longitudinal component of the end-to-end vector
showcases strong nonlinear effects in the polymer dynamics, and we
identify an effective $t^{7/8}$ power-law regime in its
time-dependence. We show that the nonlinear effects in the MSD of the
longitudinal component of the end-to-end vector increase with
increasing length for dsDNA; nevertheless, in comparison to the full
mean-square displacement of the end-to-end vector the nonlinear
effects remain small at all times. It is in this sense we state that
the linearized dynamics suffices for dsDNA fragments that are shorter
than or comparable to the persistence length.

From the results above, which are expanded further in the paper in
great detail, we hope that confidence can build up on our model. The
results are trivially extended to other semiflexible polymers as long
as the model parameter values are matched to the experimental ones, as
demonstrated in paper I.

The structure of this paper is as follows. Given that the model was
described in a fair amount of detail in paper I, we first briefly
introduce the model and describe its key characteristics (specially
those that are needed for the calculations presented in this paper) in
Sec. \ref{sec2}. In Sec. \ref{sec3} we elaborate on the (linearized)
polymer dynamics, and in Sec. \ref{sec4} we analytically determine
(and verify by simulations) the behavior of the autocorrelation
function of the end-to-end vector of the chain, that of a bond vector
at the middle of the chain, and the MSD of a tagged bead in terms of
the mode sums. We dedicate Sec. \ref{sec5} to separating the
longitudinal and transverse components of dynamical quantities, and
Sec. \ref{sec6} to nonlinear aspects of the full polymer dynamics,
specially the MSD of the longitudinal component of the end-to-end
vector. We finish the paper in Sec. \ref{sec7} with a comparison of
our results with those of the WLC model.

\section{The model, the groundstate of the Hamiltonian, and its
mechanical properties in the Hessian approximation\label{sec2}}

In this section we briefly introduce the model, the groundstate of the
Hamiltonian, and its key characteristics in the Hessian approximation.

\subsection{The model\label{sec2a}}

We describe the polymer chain consisting of $(N+1)$ beads, located at
${\bf r}_0,\ldots,{\bf r}_N$ by the Hamiltonian (with stretching,
bending and length parameters $\lambda$ and $\kappa$ and $d$
respectively)
\begin{eqnarray} {\cal H}=\lambda \sum^N_{n=1} (|{\bf u}_n|-\!d)^2- 2
\kappa \sum^{N-1}_{n=1} {\bf u}_n \cdot {\bf u}_{n+1}.
\label{e1}
\end{eqnarray} Here ${\bf u}_n$ is the bond vector between the
$(n-1)$-th and the $n$-th beads. Given that in the limit of nonzero
$d$ and large $\lambda$ and $\kappa$, it corresponds to the WLC, it is
useful to take $\nu=\kappa/\lambda$ as a parameter of the
model. Hereafter, with dimensionless temperature $T^*=k_BT/(\lambda
d^2)$, and having made ${\bf u}_n$'s unit of length, we write the
Hamiltonian as
\begin{eqnarray} \frac{\cal
H}{k_BT}\!=\!\frac{1}{T^*}\!\left[\sum^N_{n=1} (|{\bf
u}_n|\!-\!1)^2\!-\!  2 \nu\!\! \sum^{N-1}_{n=1}\! {\bf u}_n\!  \cdot\!
{\bf u}_{n+1}\! \right]\!\!.
\label{e2}
\end{eqnarray} Stability of the Hamiltonian naturally requires
$\nu<1/2$. The two parameters for this model, $T^*$ and $\nu$ can be
fixed by matching to the force-extension curve. This was carried out
for dsDNA in paper I, leading to the values $T^*=0.034$, $\nu=0.35$,
and correspondingly, the persistence length $l_p=(\nu/T^*)bd=114bd$,
where the value of $b$ is the equilibrium bond length in dimensionless
units, defined in Eq.~(\ref{e4}). The quantity $bd$ corresponds to the
length of a dsDNA basepair $\approx3$\AA\/ \cite{hans}. We will use
these values for our simulations all throughout the paper, while we
stress that the simulation results are trivially extended for other
semiflexible polymers as long as the parameters $T^*$ and $\nu$ are
adjusted to match the corresponding force-extension curve.

Note from Eq.~(\ref{e1}) that the strength of the stretching term is
given by $\lambda d^2=k_BT/T^*\approx30k_BT$. This means that for all
practical purposes the bond lengths of the chain are fluctuating
around $bd\approx3$\AA\/ only by a few percent.

\subsection{Groundstate of the Hamiltonian\label{sec2b}}

The groundstate of the Hamiltonian is obtained from the condition
\begin{eqnarray} \frac{\partial[{\cal H}/(k_BT)]}{\partial {\bf u}_n
}= \frac{1}{T^*}[ {\bf u}_n\!-\!\hat{\bf u}_n\!-\!\nu ({\bf
u}_{n+1}\!+\!{\bf u}_{n-1} )]=0.
\label{e3}
\end{eqnarray} In the groundstate all the bond vectors align --- i.e.,
in the groundstate the chain configuration is that of a straight rod.
An exact expression can be found for the bond length \cite{hans}
\begin{eqnarray} |{\bf u}_n| = \frac{1}{1-2\nu} \left[
1-\frac{\cosh\{\alpha(N+1-2n)\}}{\cosh\{\alpha(N+1)\}} \right],
  \label{e3a}
\end{eqnarray} with $\alpha=\frac{1}{2} \cosh^{-1}(1/(2 \nu))$. Far
away from the chain ends the bond lengths become equal [$=b$],
satisfying \cite{hans}
\begin{eqnarray} b-1-2\nu b=0,\,\,{\rm or}\quad b=1/(1-2\nu).
\label{e4}
\end{eqnarray} Note the value $bd=d/(1-2\nu)$ corresponds to the
length of a dsDNA basepair $\approx3$\AA\/, which determines the
choice for $d$ in case of dsDNA \cite{hans}.

\subsection{Mechanical properties of the model: the Hessian
approximation\label{sec2c}}

The energy around the minimum (i.e., around the groundstate of the
Hamiltonian) varies quadratically with the $\{{\mathbf r}_n\}$s at the
first order of approximation, and is dictated by the Hessian matrix
$\partial^2{\cal H}/(\partial{\mathbf r}_m\partial{\mathbf r}_n)$.

The eigenvectors of the Hessian matrix have three branches: a
longitudinal one, for which the eigenvectors are aligned along the
groundstate configuration (straight rod) of the chain, and two
identical sets of transverse ones. The decay spectrum of the
longitudinal modes is given by \cite{hans}
\begin{eqnarray} \zeta^l_p=2 \left[1- \cos \left(\frac{p\pi}{
N+1}\right)\right]
\left[1-2\nu\cos\left(\frac{p\pi}{N+1}\right)\right],
\label{m2}
\end{eqnarray} which increase as $(p/N)^2$ for low $p$-values, with
$p=0,1,2, \cdots, N$. The modes $p=0$, having a zero eigenvalue,
correspond to the center-of-mass motion.  The transverse modes have
also a zero eigenvalue for $p=1$, corresponding to the invariance of
the Hamiltonian to an overall transverse rotation of the
chain. Further, the transverse eigenspectrum agrees very well with the
approximate expression \cite{hans}
\begin{eqnarray} \zeta^t_p \approx 4\nu \left(1-\cos
\left[\frac{p\pi}{N+1}\right]\right)
\left(1-\cos\left[\frac{(p-1)\pi}{N+1}\right]\right),
\label{m3}
\end{eqnarray} i.e., $\zeta^t_p\sim p^2 (p-1)^2/N^4$ for small
$p$. For low $p$ this behavior is characteristic for semiflexible
chains \cite{farge}.

The eigenfunctions of the longitudinal modes are the same as those for
the Rouse modes:
\begin{eqnarray} \phi_{n,p} = \left(\frac{2}{N+1}\right)^{1/2} \cos
\left[\frac{p(n+1/2)\pi}{N+1}\right].
\label{e13}
\end{eqnarray} The transverse mode eigenfunctions are not markedly
different, but have to be determined numerically. Note also that the
eigenfunctions (\ref{e13}) are even for even $p$ under reversal of the
bead indices [i.e., $i\leftrightarrow(N-i)$] and the odd $p$ are odd.

The motion of the center-of-mass is independent of the other modes of
the system and it plays no role for the properties that we consider,
except for the MSD of the tagged bead, which is partly due to internal
motion and partly driven by the center-of-mass motion. For the other
properties it is convenient to view the chain in the coordinate system
where the center-of-mass is at the origin of the coordinate system.

{\it The modes form a complete basis}, so any quantity that is a
linear expression in the coordinates, can be expressed in terms of
this basis. This is the key to analytically evaluate the quantities in
the next section.  The transverse modes corresponding to $p=1$ are
special in the sense that they induce a rotation of the reference
groundstate.  The eigenvector has therefore the form
\begin{eqnarray} \phi^t_{n,1} = - r^{(0)}_n/ \sqrt{I}, \quad \quad
{\rm with} \quad \quad I=\sum_n (r^{(0)}_n)^2=b^2 N^3/12.
\label{e8a}
\end{eqnarray} $I$ is the moment of inertia of the groundstate in
dimensionless units \cite{hans}.  Actually the representation of the
configuration in terms of modes with respect to a groundstate is
redundant. One could take any fixed groundstate as reference.  As the
two transverse modes $p=1$ do not decay, they would grow in size due
to the random forces as random walkers, indicating that the reference
groundstate is not anymore in line with the actual shape of the
chain. However by rotating the direction of the reference groundstate
one can set the two transverse modes for $p=1$ equal to zero. The
representation with vanishing transverse modes $p=1$ is unique.

\section{Polymer dynamics\label{sec3}}

Although the concept of the modes originate from the Hessian
approximation, they can be used to describe the {\it full\/} polymer
dynamics, since they provide a {\it complete\/} set of orthogonal
basis functions. In this section we show how this can be achieved.

If we denote the modes by ${\mathbf \Psi}_p(t)$, then they can be
expressed in terms of the bead co-ordinates $\{{\bf r}\}_n(t)$ as
\begin{eqnarray} {\mathbf \Psi}_p(t)=\sum_n \,[{\mathbf
r}_n(t)-{\mathbf r}^{(0)}_n(t)] \, \phi_{n,p},
\label{e8}
\end{eqnarray} where ${\mathbf r}^{(0)}_n(t)$ is the position of the
$n$-th bead in the reference groundstate. It is important to note here
that the instantaneous position of the center-of-mass of the reference
groundstate, which is also its midpoint, coincides with the
instantaneous position of the center-of-mass of the chain itself.
Here $\phi_{n,p}$ symbolically stands for both the longitudinal as
well as the transverse eigenfunctions. As the transformation from
${\mathbf r}_n(t)-{\mathbf r}^{(0)}_n(t)$ to ${\mathbf \Psi}_p(t)$ is
orthogonal, the inverse relation reads
\begin{eqnarray} {\mathbf r}_n(t)={\mathbf r}^{(0)}_n(t)+\sum_p
\phi_{n,p} \, {\mathbf \Psi}_p(t),
\label{e9}
\end{eqnarray} Note that the modes give {\it deviations\/} from the
reference groundstate.  Thus, although the longitudinal modes have the
same eigenfunction and eigenvalue as the Rouse modes,
${\mathbf\Psi}_p(t)$ is not the same as the Rouse mode amplitude,
since the latter is expressed in terms of the positions ${\mathbf
r}_n(t)$, while the former are expressed in terms of the deviations
${\mathbf r}_n(t)-{\mathbf r}^{(0)}_n(t)$.

In the overdamped limit the dynamical equation of the $n$-th bead is
given by
\begin{eqnarray} \frac{d{\bf r}_n}{dt}=-\xi^{-1}\frac{\partial {\cal
H}}{\partial {\bf r}_n}+{\bf g}_n(t),
\label{hesseq}
\end{eqnarray} where $\xi$ is the friction coefficient acting on the
bead due to the viscosity of the surrounding medium in the overdamped
description, and ${\bf g}_n(t)$ is the thermal noise term satisfying
the fluctuation-dissipation relation $\langle{\bf g}_m(t){\bf
g}_n(t')\rangle=(2k_BT/\xi){\bf I}\delta_{mn}\delta(t-t')$ with ${\bf
I}$ as the identity tensor. Since the eigenmodes of the Hessian matrix
provide a complete orthogonal basis, Eq.~(\ref{hesseq}) can simply be
rewritten in terms of the mode amplitudes $\{{\bf\Psi}_p\}$. Further,
in terms of the dimensionless time unit $\tau=\lambda t/\xi$ their
dynamical evolution of the modes is given by the Langevin equation
\begin{eqnarray} \frac{d{\mathbf \Psi}_p(\tau)}{d\tau} =
-\zeta_p\,{\mathbf \Psi}_p(\tau)+ {\mathbf H}_p (\tau)+{\mathbf
G}_p(\tau),
\label{e10}
\end{eqnarray} where $\zeta_p$ is used to collectively denote the
decay constants for the modes ($\zeta_p=\zeta^l_p$ for longitudinal
and $=\zeta^t_p$ for transverse modes) and ${\mathbf G}_p$ are random
forces obeying the fluctuation-dissipation theorem
\begin{eqnarray} \langle{\mathbf G}_p(\tau) {\mathbf
G}_q(\tau')\rangle=2T^*{\mathbf I}\delta_{pq}\delta(\tau-\tau').
\label{e11}
\end{eqnarray} The term ${\mathbf H}_p (\tau)$ represents the coupling
between the modes.  The Hessian matrix diagonalizes the Hamiltonian to
second order in the deviations from the reference groundstate. The
term ${\mathbf H}_p (\tau)$, henceforth referred to as the ``coupling
force'', results from higher order deviations.  The average magnitude
of the modes in the Hessian approximation of the Hamiltonian is of
order $\sqrt{T^*}$ as the random forces ${\mathbf G}_p$ have this
magnitude according to Eq. (\ref{e11}).  The coupling forces are of
order $T^*$, since they result from products of modes. In the next
section we leave them out, and this defines precisely the linearized
regime, where thermal fluctuations are important, but the modes remain
the independent degrees of freedom for the dynamics.

\subsection{Linearized Dynamics \label{sec3a}}

We now outline the simplifications if one neglects the coupling force
${\bf H}_p$.  Omitting the coupling force leads to modes evolving
independently in time according to an Ornstein-Uhlenbeck process. For
such a process, the conditional probability of having a value
${\mathbf \Psi}_p (\tau)$, given that it had the value ${\mathbf
\Psi}_p(0)$ at $\tau=0$, follows as
\begin{eqnarray} P[{\mathbf \Psi}_p (\tau)|{\mathbf \Psi}_p(0)] =
\frac{1}{w_p(\tau) \sqrt\pi} \exp \left[ - \frac{(\Delta {\mathbf
\Psi})^2}{2 w^2_p(\tau)}\right],
\label{e26}
\end{eqnarray} with
\begin{eqnarray} \Delta {\mathbf \Psi} = {\mathbf \Psi}_p
(\tau)-{\mathbf \Psi}_p(0) \exp(-\zeta_p \tau),
\label{e27}
\end{eqnarray} where $\zeta_p$ is used to collectively denote
$\zeta^l_p$ or $\zeta^t_p$ as applicable. In other words, the
conditional average of ${\mathbf \Psi}_p (\tau)$, given the value
${\mathbf \Psi}_p(0)$ at $\tau=0$, equals
\begin{eqnarray} \langle {\mathbf \Psi}_p (\tau) \rangle_{{\mathbf
\Psi}_p(0)} = {\mathbf \Psi}_p(0) \exp(-\zeta_p \tau).
\label{e28}
\end{eqnarray} The width $w_p (\tau)$ of the distribution is given by
\begin{equation} w^2_p (\tau) = T^* \frac{1- \exp(-2\zeta_p
\tau)}{\zeta_p}.
\label{e29} \end{equation} Knowing the temporal evolution, averages
can be worked out by using the equilibrium averages
\begin{equation} \label{e29c} \langle \mathbf{\Psi}_p \rangle = {\bf
0}, \quad \quad \langle \mathbf{\Psi}_p \mathbf{\Psi}_q \rangle =
\delta_{p,q}\, {\bf I}\,\frac{T^*}{\zeta_p}.
\end{equation}

With the rules (\ref{e28}) and (\ref{e29c}) the equilibrium averages
of time-dependent correlations functions can be easily evaluated. We
give, as example, the end-to-end vector ${\bf L}$, defined as the
difference between the first and last bead of the chain
\begin{equation} \label{e29b} {\bf L}(\tau) = {\bf r}_N (\tau) - {\bf
r}_0 (\tau).
\end{equation} It can be written as a sum of three vectors
\begin{equation} \label{e19} {\bf L}(\tau) = {\bf r}_N (\tau) - {\bf
r}_0 (\tau) = {\bf L}^{(0)}(\tau) + {\bf L}^l(\tau) + {\bf L}^t(\tau).
\end{equation} The first ${\bf L}^{(0)}(\tau)$ is the contribution of
the reference groundstate
\begin{equation} \label{e19a} {\bf L}^{(0)}(\tau) = L^{(0)} \hat{\bf
e}_0 (\tau),
\end{equation} where $\hat{\bf e}_0 (\tau)$ is the orientation of the
reference groundstate.  The second $ {\bf L}^l(\tau)$ is the
contribution of the longitudinal modes given by
\begin{equation} \label{e19b} {\bf L}^l(\tau) = \sum_p L^l_p \, {\bf
\Psi}^l_p (\tau).
\end{equation} The coefficient $L^l_p$ is the coupling of the
end-to-end vector to the longitudinal modes
\begin{equation} \label{e20} L^l_p=\phi^l_{N,p} - \phi^l_{0,p} \simeq
\left( \frac{2}{N}\right)^{1/2} [(-1)^p-1].
\end{equation} The third (and the last) contribution in (\ref{e19}) is
the sum over the transverse modes
\begin{equation} \label{e20a} {\bf L}^t(\tau) = \sum_p L^t_p \, {\bf
\Psi}^t_p (\tau),
\end{equation} with the transverse coupling coefficients $L^t_p$
\begin{equation} \label{e20b} L^t_p=\phi^t_{N,p}-\phi^t_{0,p}.
\end{equation} The temporal evolution of the longitudinal and
transverse parts is implied by that of the corresponding modes as
given by Eq.  (\ref{e28}). The orientation of the reference
groundstate is implicitly given by the requirement that the two
transverse components of the first mode $p=1$ remain absent. Its
evolution is purely diffusive and independent of the mode
evolution. In the independent mode approximation one finds
\begin{equation} \label{e24} \langle \hat{\bf e}_0 (\tau) \cdot
\hat{\bf e}_0 (0) \rangle=\exp(-2 D_r \tau) \quad \quad {\rm with}
\quad \quad D_r=\frac{T^*}{I},
\end{equation} where $D_r$ is the rotational diffusion coefficient.

\section{Analytical expressions for some interesting observables in
the linearized dynamics, and comparison with $\mbox{ds}$DNA
simulations\label{sec4}}

In this section we give the expressions for the decay of the
correlation function of a number of quantities. We provide analytical
expressions in the linearized dynamics approximation, as they follow
from the corresponding modes, and compare the analytical results to
the simulations. The simulations, incorporating the {\it full\/}
dynamics, consist of a simple time-forward integration scheme for
Eqs. (\ref{e10}-\ref{e11}).

\subsection{The end-to-end vector autocorrelation
function\label{sec4a}}

We start with the end-to-end vector autocorrelation function
\begin{equation} \label{e23a} C_L(\tau) = \langle {\bf L} (\tau) \cdot
{\bf L} (0) \rangle,
\end{equation} and evaluate this quantity analytically in the
linearized approximation.
 
The decay of the modes is independent of the orientation of the
groundstate. In the linearized approximation the correlation functions
are a product of the correlation function Eq. (\ref{e24}) of the
orientation of the groundstate and the correlation function of the
modes. Since, according to Eq. (\ref{e29c}), the equilibrium average
of a single mode vanishes and transverse and longitudinal are
orthogonal, we get for the correlation function (\ref{e23a}) the sum
of three contributions
\begin{equation} \label{e23c} C_L(\tau) = \left( [L^{(0)}]^2 + C^l_L
(\tau) + C^t_L (\tau) \right) \exp(-2 D_r \tau),
\end{equation} where the contributions of the longitudinal and
transverse mode-sums are given by
\begin{equation} \label{e31} C^l_L(\tau)= T^* \, \sum_{p=1}
\frac{[L^l_p]^2 \exp (-\zeta^l_p \tau) }{\zeta^l_p}, \quad \quad \quad
C^t_L(\tau)= 2 T^* \, \sum_{p=2} \frac{[L^t_p]^2 \exp (-\zeta^t_p
\tau) }{\zeta^t_p},
\end{equation} In the transverse sum the mode $p=1$ is excluded since
it is eliminated.  The numerical evaluation of these mode sums is
straightforward.  Regarding the behavior of these mode-sums we
distinguish three regimes in time.
\begin{itemize}
\item For times where all the exponents $\zeta_p \tau $ are small, the
exponential may be expanded, leading to a power series in $\tau$. Note
that to linear order in $\tau$, each mode equally contributes. This
regime extends to times of order 1, since the highest modes decay with
a coefficient of order 1.
\item For longer times the higher modes gradually start to drop out of
the summation.  This intermediate regime is actually the most
interesting, since the sum over modes still contains many smoothly
varying terms in which the low-$p$ modes have the largest influence.
In this regime the sums may be replaced by integrals of which the
asymptotic properties are analyzed in Appendix \ref{stret}, leading to
a time dependence in terms of fractional powers of $\tau$. The regime
extends to a characteristic time $\tau^*$, which is of the order of
$N^4\sim (\zeta^t_2)^{-1}$ for the transverse modes and of order $N^2
\sim (\zeta^l_1)^{-1}$ for the longitudinal modes.  It means that the
window, in which typical longitudinal effects can be seen, is small
with respect to that of the transverse effects.  Moreover the
transverse modes overshadow the longitudinal modes by a factor $N^2$,
due to the decay constant $\zeta^t_p$ in the denominator which is, for
low $p$, smaller than $\zeta^l_p$ by a factor $N^2$. In the
intermediate regime the decay, due to reorientation of the
groundstate, is still small.
\item For $\tau >\tau^*$ all the exponents are large and the
exponentials small. Thus what remains is the decay due to rotational
reorientation, governed by the diffusion constant $D_r$. So only
exponential decay is observed in the correlation functions.
\end{itemize}
\begin{figure}[h]
\begin{minipage}{0.48\linewidth}
\includegraphics[width=\linewidth]{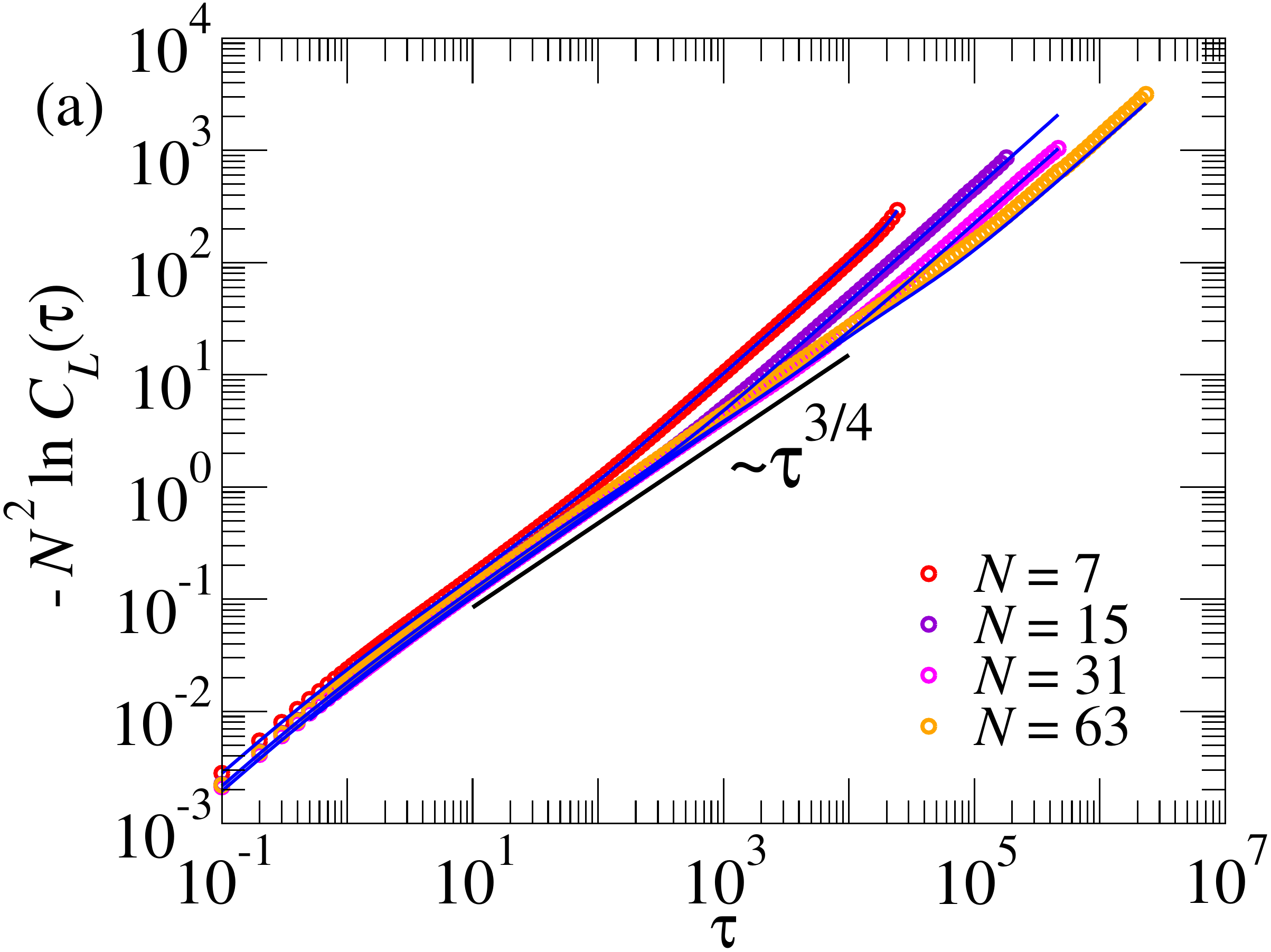}
\end{minipage} \hspace{5mm}
\begin{minipage}{0.46\linewidth}
\includegraphics[width=\linewidth]{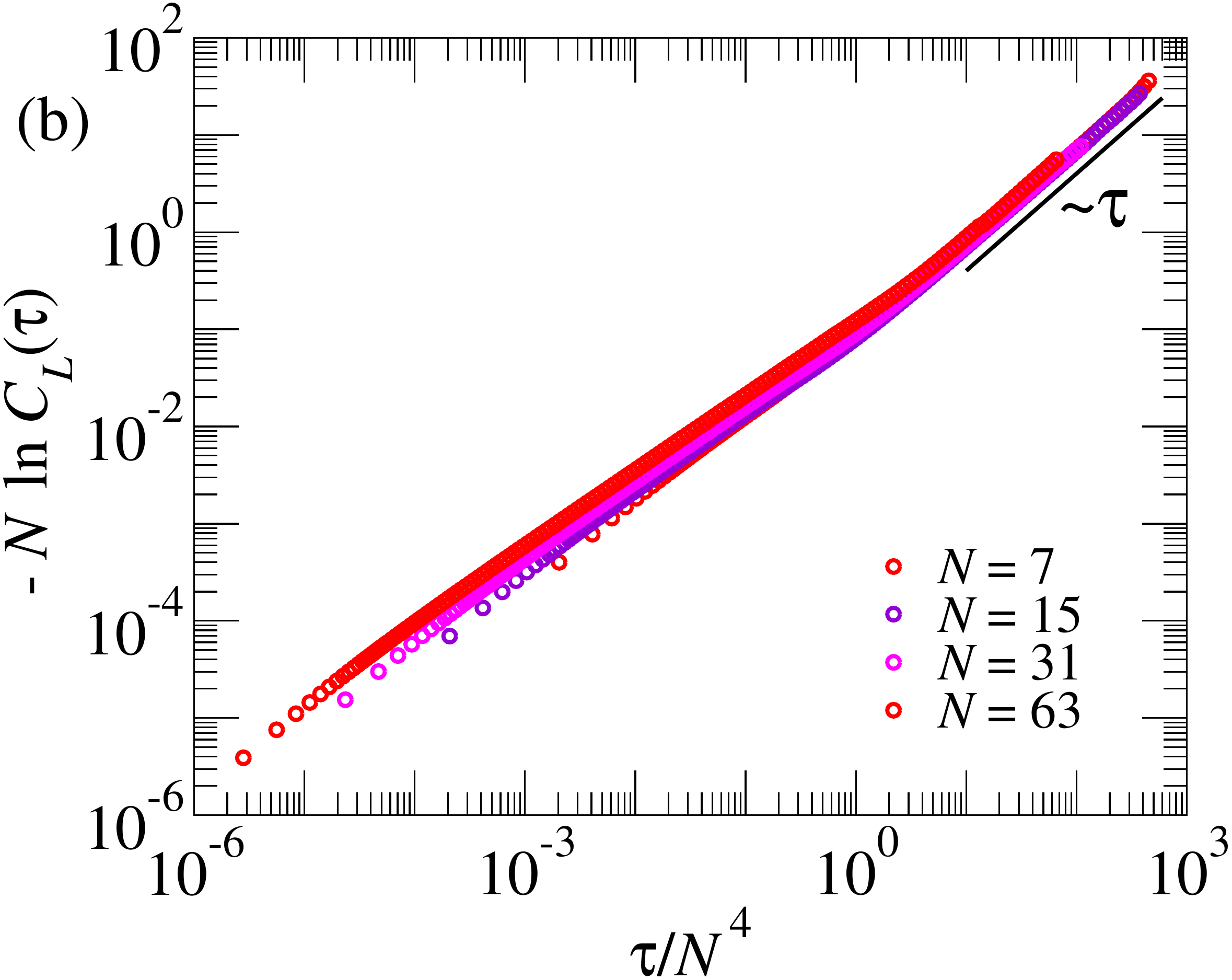}
\end{minipage}
\caption{(a) At intermediate times the end-to-end autocorrelation
function $C_L(\tau)$ exhibits stretched exponential behavior with
exponent $\tau^{3/4}/N^2$. Note that the persistence length of dsDNA
corresponds to $N=114$ in dimensionless units. The simulation data are
represented by points, while the solid blue lines, in excellent
agreement with the simulation data, represent the corresponding the
combined effect of the mode-sums (\ref{e31}). Note also that we cannot
reliably use the Hessian approximation of the Hamiltonian too much
beyond $N=63$. (b) The stretched exponential behavior lasts till the
terminal time $\tau^*\sim N^4$, beyond which the decay becomes
exponential in $\tau/N^3$. \label{fig1}}
\end{figure}

We now discuss the analytical behavior of the correlation function in
the intermediate time regime and focus on the transverse modes, being
the most interesting.  The deviation of the correlation from its
initial value reads
\begin{eqnarray} 1 - \frac{C_L (\tau)}{C_L (0)} \simeq \frac {16
T^*}{b^2 N^3} \sum_{p=3,5, \cdots} \frac{1}{\zeta^t_p} [1-
\exp(-\zeta^t_p \tau)].
\label{e35}
\end{eqnarray} Sums of the type (\ref{e35}) are typical for the
correlation functions that we consider. They are worked out in
Appendix \ref{stret}.  With the function $F_0 (\tau)$ defined in
Eq.~(\ref{A1}) and the result (\ref{A5}) we get
\begin{eqnarray} 1 - \frac{C_L (\tau)}{C_L(0)} \simeq \frac{16
T^*}{b^2 N^3} F_0 (\tau) \simeq \Gamma(1/4)\frac{8 T^* \tau^{3/4} }{3
\pi b^2 N^2 \nu^{1/4}}.
\label{e36}
\end{eqnarray} So at intermediate times, where the deviations of the
initial value are still small, the correlation function decays in time
as a stretched exponential:
\begin{eqnarray} C_L(\tau) \simeq C_L(0) \,\exp\left[- \Gamma(1/4)
\frac{8 T^* \tau^{3/4} } {3 \pi b^2 N^2 \nu^{1/4}} \right].
\label{e37}
\end{eqnarray}

We note that in the exponent of (\ref{e37}), combination $T^*
\tau^{3/4}/N^2$ can be written as the ratio
$(N/l_p)[\tau^{1/4}/N]^3$. Since $N/l_p$ is of order 1 in the region
of our interest, the ratio is a function of $[\tau/\tau^*]^3$ with
$\tau^* =N^4$, which is of the order of the slowest transverse decay
time $1/\zeta^t_2$.  So the stretched exponential behavior crosses
over to exponential decay after time $\tau^*$. Note that, due to
numerical factors, $\tau^*$ is still small in comparison to $1/D_r$,
which scales as $N^3/T^*$.

As we corroborate the stretched exponential behavior of $C_L(\tau)$ by
means of direct simulations in Fig.~\ref{fig1}, we note that apart
from the mode sum (\ref{e27}) the stretched exponential behavior can
be obtained by the following simple argument. First,
$C_L(\tau)=1-\langle[{\mathbf L}(\tau)-{\mathbf
L}(0)]^2\rangle/(2\langle L^2(0)\rangle) \sim1-\langle[{\mathbf
L}(\tau)-{\mathbf L}(0)]^2\rangle/(2N^2)$ is an identity. Next, in
time $\tau_0\sim1$, independent of its length, the chain ends move in
real space by order $N^0$; i.e., $\langle[{\mathbf L}(\tau_0)-{\mathbf
L}(0)]^2\rangle\sim N^0$, and consequently, $-\log
C_L(\tau_0)\sim1/N^2$. Further, beyond time $\tau^*\sim N^4$ the
correlation function decays exponentially due to rotational diffusion
dynamics with diffusion coefficient $D_r\sim N^{-3}$ leading to $-\log
C_L(\tau^*)\sim N$.  Thus, if one assumes that the $ -\log C_L(\tau)$
values at $\tau_0\sim 1$ and at $\tau^*\sim N^4$ are bridged by an
exponential function of a single character, then the only solution is
a stretched exponential with exponent $\tau^{3/4}/N^2$. These
arguments are confirmed in Fig. \ref{fig1}.

\subsection{The autocorrelation function for the middle bond
vector\label{sec4b}}

We now evaluate the autocorrelation function for the middle bond
vector analytically in the linearized approximation.

For the autocorrelation function for the middle bond vector we
consider a chain with an even number of beads, i.e., $N$ is odd. Like
the case of the end-to-end vector, the middle bond vector ${\mathbf
u}_m(\tau)$ is described as
\begin{eqnarray} {\mathbf u}_m(\tau) = {\mathbf u}_m^{(0)} (\tau) +
\sum_p u_p {\mathbf \Psi}_p(\tau)={\mathbf u}_m^{(0)} (\tau) +
\sum_p\left[\phi_p^{(N+1)/2} - \phi_p^{(N-1)/2}\right] {\mathbf
\Psi}_p (\tau).
\label{e29a}
\end{eqnarray} Here ${\mathbf u}_m^{(0)} (\tau)$ is the middle bond
vector for the reference groundstate configuration (straight
rod). Once again, ${\mathbf u}_m$ is odd under reversal of the
numbering of the beads, so the even modes do not contribute to the sum
(\ref{e29a}). Comparing Eqs. (\ref{e13}) and (\ref{e8}) we get
\begin{eqnarray} u_p = \phi^{(N+1)/2}_p -\phi^{(N-1)/2}_p \simeq -2
\left(\frac{2}{N+1}\right)^{1/2} \sin\left(\frac{p\pi}{2}\right)\sin
\left[\frac{p\pi}{2(N+1)} \right],
\label{e30a}
\end{eqnarray} where the second identity in Eq. (\ref{e30a}) is strict
for the longitudinal components and approximate for the transverse
ones.

Beyond the time $\tau^*$, since the vector ${\mathbf u}_m^{(0)}(\tau)$
in the reference groundstate undergoes the same rotational diffusion,
the expression for the autocorrelation function for the middle bond
vector $C_m(\tau)=\langle{\mathbf u}_m(\tau)\cdot{\mathbf
u}_m(0)\rangle/\langle u^2_m(0)\rangle$ is similar to those for the
end-to-end vector, allowing us to write
\begin{eqnarray} C_m(\tau)=\frac{1+ \sum_{p\in\text{odd}} \, [u^2_p\,
T^*] / [(u^{(0)}_m)^2 \zeta_p] \, \exp(-\zeta_p
\tau)}{1+\sum_{p\in\text{odd}} \, [u^2_p\, T^*] / [(u^{(0)}_m)^2
\zeta_p]}\exp(-2 D_r \tau),
\label{e31b}
\end{eqnarray} wherein, once again, the sum does not include $p=1$
transverse modes, and each value of $p$ for the transverse modes needs
to be counted twice.

The analysis of the sum over the slow modes is given in the Appendix.
With the function $F_2 (\tau)$ defined in Eq.~(\ref{A1}) and the
result (\ref{A6}) we get
\begin{eqnarray} C_m(0) - C_m (\tau) \simeq \frac {\pi^2T^*}{3 I} F_2
(\tau) \simeq \Gamma(3/4)\frac{N^3 T^* \tau^{1/4} }{6 \pi I
\nu^{3/4}}.
\label{e31a}
\end{eqnarray} This gives a stretched exponential
\begin{eqnarray} C_m(\tau) \simeq \exp\left[-\Gamma(3/4) \frac{T^* N^3
\tau^{1/4}}{6 \pi I \nu^{3/4}} \right],
\label{e32a}
\end{eqnarray} In order to appreciate the time regime in which this
stretched exponent features we note again that the factor $T^*$
effectively counts as a factor $N^{-1}$ and that the exponential
behavior lasts up to $\tau^*\sim N^4$. In that time span the exponent
is of order unity and after $\tau^*$ the correlation follows the
diffusive behavior of the end-to-end vector.

\begin{figure}[h]
\begin{minipage}{0.48\linewidth}
\includegraphics[width=\linewidth]{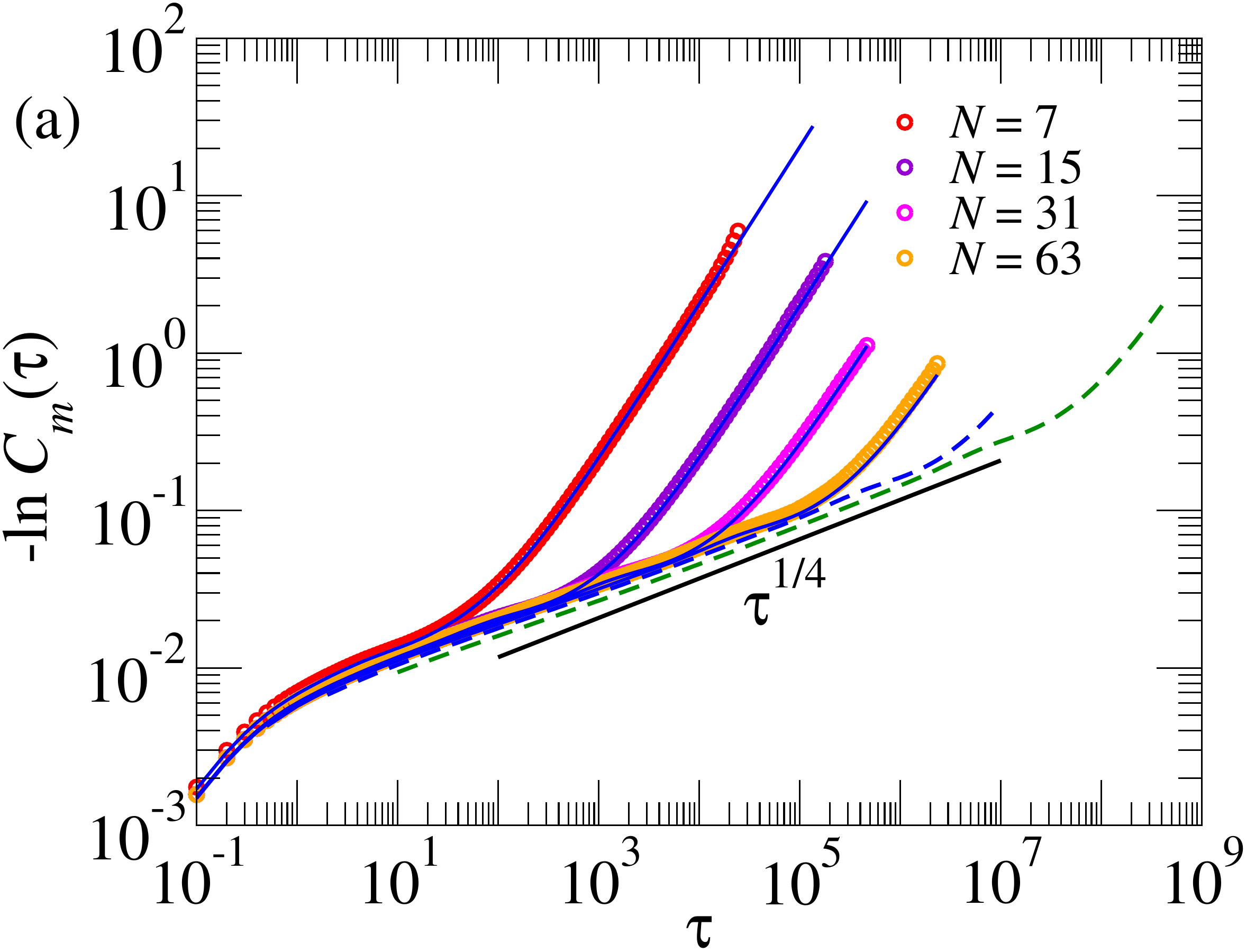}
\end{minipage} \hspace{5mm}
\begin{minipage}{0.46\linewidth}
\includegraphics[width=\linewidth]{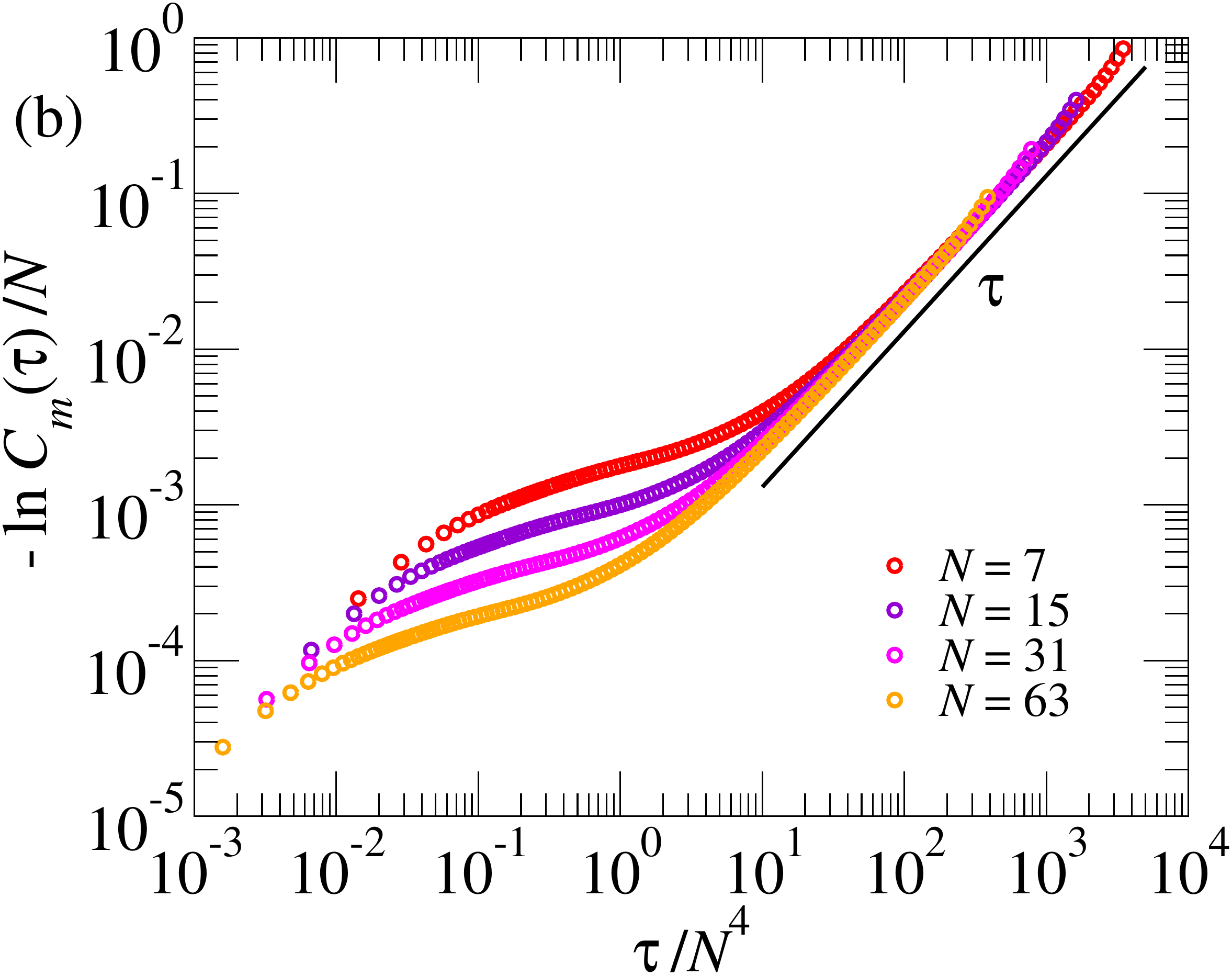}
\end{minipage}
\caption{(a) At intermediate times the end-to-end autocorrelation
function $C_m(\tau)$ exhibits stretched exponential behavior with
exponent $\tau^{1/4}$. The simulation data are represented by points,
while the solid blue lines, in excellent agreement with the simulation
data, represent the corresponding mode sums (\ref{e31b}). The blue and
green dashed lines correspond to the mode sum (\ref{e31b}) for $N=127$
and $N=255$ respectively. (b) The stretched exponential behavior lasts
till the terminal time $\tau^*\sim N^4$, beyond which the decay
becomes exponential in $\tau/N^3$. \label{fig2}}
\end{figure}

The verification of this result can be found in Fig. \ref{fig2}, which
also demonstrates that the mode sum (\ref{e32a}) exhibits strong
finite-size effects, i.e. the sum and the asymptotic fractional power
differ substantially.  This is purely an issue related to the sum
(\ref{e31b}) --- the sum can be carried out for indefinitely large
values of $N$ (but $N$ cannot be indefinitely increased in
simulations, as, given that $l_p=114$ in dimensionless units, we
cannot reliably use the Hessian approximation of the Hamiltonian too
much beyond $N=63$) --- and as can be seen in Fig.~\ref{fig2}, with
increasing $N$ the mode sum (\ref{e31b}) does indeed approach the
expected stretched exponential behavior (\ref{e32a}).

Like in the case of $C_L(\tau)$, the stretched exponential behavior
for $C_m(\tau)$ between $\tau\sim O(1)$ and $\tau=\tau^*$ can also be
argued from the real-space mean-square displacement of the middle bead
in the following manner. First, $C_m(\tau)=1-2\langle[{\mathbf
u}_m(\tau)-{\mathbf u}_m(0)]^2\rangle/\langle
u_m^2(0)\rangle\sim1-2\langle[{\mathbf u}_m(\tau)-{\mathbf
u}_m(0)]^2\rangle/N^0$ is an identity. Next, in time $\tau_0\sim1$ the
two beads connecting the middle bond vector move in real space by
order $N^0$; i.e., $\langle[{\mathbf u}_m(\tau_0)-{\mathbf
u}_m(0)]^2\rangle\sim N^0$, and consequently, $-\log C_m(\tau_0)\sim
N^0$. Further, beyond time $\tau^*\sim N^4$ the middle bond vector
must undergo rotational diffusion with diffusion coefficient $D_r\sim
N^{-3}$, leading to the result $-\log C_m(\tau^*)\sim N$. Thus, if one
assumes that $-\log C_m(\tau)$ values at $\tau_0\sim 1$ and at
$\tau^*\sim N^4$ are bridged by an exponential function of a single
character, then the only solution is a stretched exponential with
exponent $1/4$. These arguments are confirmed in Fig.~\ref{fig2}.

\subsection{The mean-square displacement of the middle
bead\label{sec4c}}

Finally, we evaluate the mean-square displacement of the middle bead
in the linearized approximation.
 
In order to avoid setting up additional simulations for chains with
odd number of beads, we continue in this section with chain with even
number of beads. We are then interested in the MSD of the
center-of-mass of the two middlemost beads; i.e., the mean ${\mathbf
r}_m(\tau)$ of the location of the $(N-1)/2$ and $(N+1)/2$-th beads,
${\mathbf r}_m(\tau) =\frac12[{\mathbf r}_{(N+1)/2}(\tau)+ {\mathbf
r}_{(N-1)/2}(\tau)]$ and express it in terms of the modes as
\begin{eqnarray} {\mathbf r}_m(\tau)\! ={\mathbf R}_{\text{cm}}(\tau)
+ \sum_p r_p {\mathbf \Psi}_p (\tau) \equiv{\mathbf
R}_{\text{cm}}(\tau) +\!\tilde{\mathbf r}_m (\tau),
\label{e33a}
\end{eqnarray} since, as noted before, the instantaneous location of
the midpoint of the reference groundstate coincides with that of the
chain's center-of-mass. The corresponding mode coefficient is then
given by
\begin{eqnarray} r_p = \phi^{(N+1)/2}_p +\phi^{(N-1)/2}_p \simeq
2\left(\frac{2}{N+1}\right)^{1/2} \cos\left(\frac{p\pi}{2}\right)\cos
\left[\frac{p\pi}{2(N+1)} \right],
\label{e34a}
\end{eqnarray} where the second identity in Eq. (\ref{e33a}) is strict
for the longitudinal components and approximate for the transverse
ones.

The MSD of the middle bead is the sum of two terms
 \begin{eqnarray} \langle\Delta r_m^2(\tau)\rangle=\langle[\Delta
\mathbf R_{\text{cm}} (\tau)]^2 \rangle + \langle [\Delta
\tilde{\mathbf r}_m(\tau)]^2 \rangle,
 \label{e34c}
\end{eqnarray} as the cross terms vanish because the center-of-mass
motion of the chain is independent of the internal motion represented
by $\tilde{\mathbf r}_m(\tau)$. The center-of-mass diffuses with
coefficient $D \sim 1/N$
 \begin{eqnarray} \langle[\Delta \mathbf R_{\text{cm}} (\tau)]^2
\rangle = 4 D \tau
 \label{e34b}
\end{eqnarray} and the internal MSD
\begin{eqnarray} \langle\Delta
\tilde{r}_m^2(\tau)\rangle=\langle[\tilde{\mathbf
r}_m(\tau)-\tilde{\mathbf r}_m(0)]^2\rangle,
\label{e35a}
\end{eqnarray} can be computed from the mode sums as above. From the
definition it is clear that $\langle\Delta \tilde{r}_m^2(\tau)\rangle$
must asymptotically approach a constant.

\begin{figure}[h]
\begin{minipage}{0.48\linewidth}
\includegraphics[width=\linewidth]{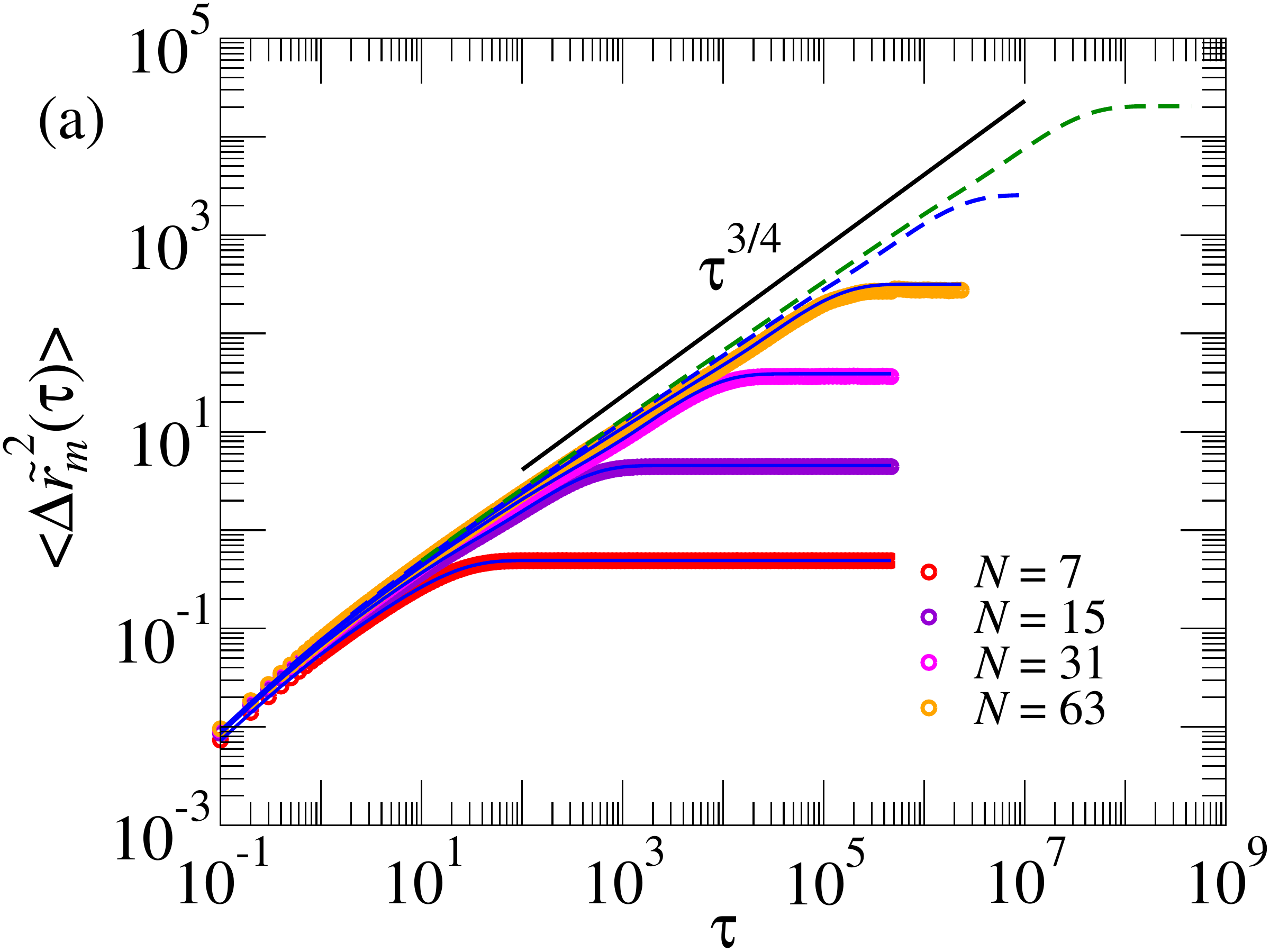}
\end{minipage} \hspace{5mm}
\begin{minipage}{0.46\linewidth}
\includegraphics[width=\linewidth]{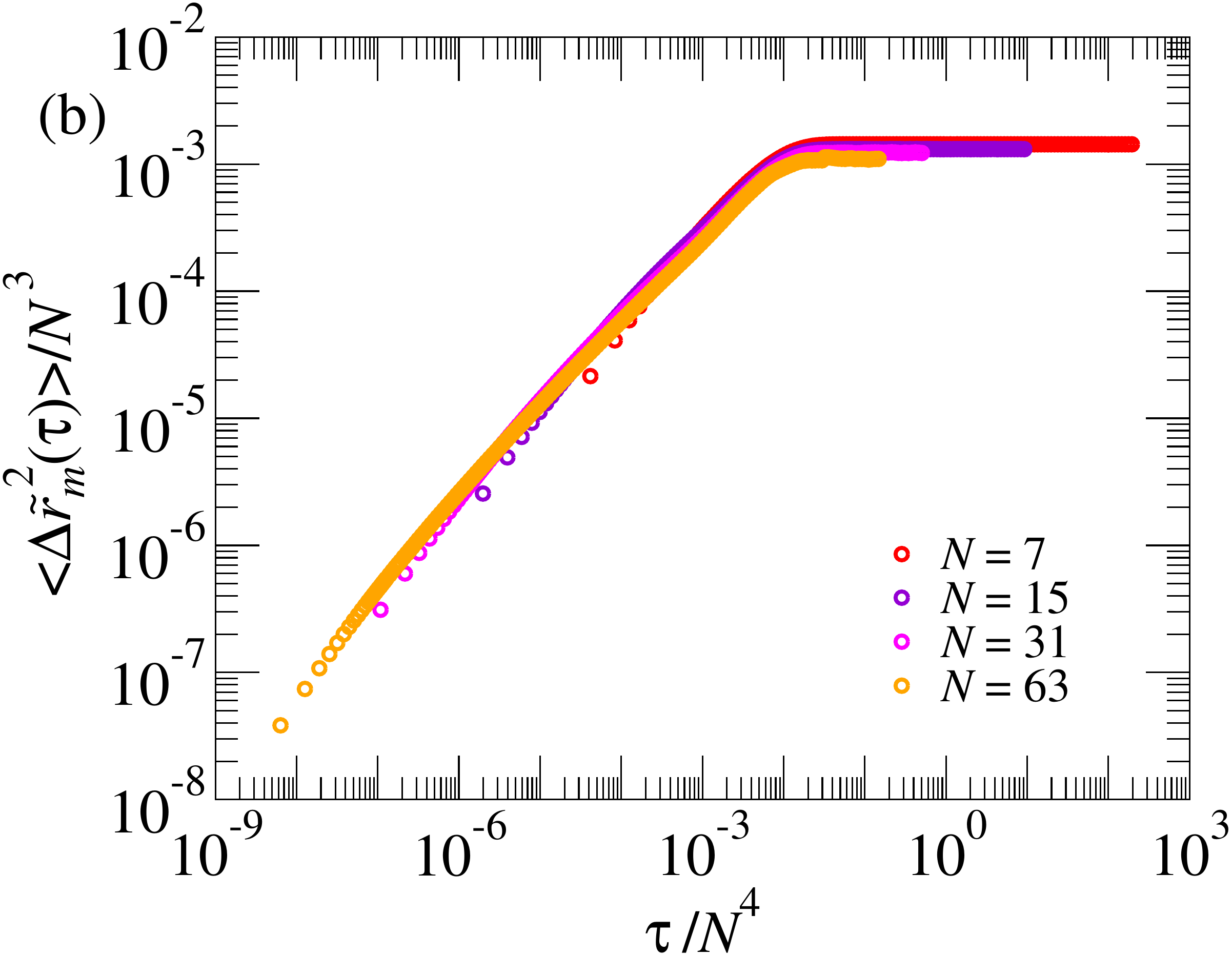}
\end{minipage}
\caption{(a) At intermediate times $\langle\Delta \tilde{r}_m^2(\tau)$
increases subdiffusively with exponent $3/4$. The simulation data are
represented by points, while the solid blue lines, in excellent
agreement with the simulation data, represent the corresponding mode
sums (\ref{e36a}). The blue and green dashed lines correspond to the
mode sum (\ref{e36a}) for $N=127$ and $N=255$ respectively. (b) The
subdiffusive behavior lasts till the terminal time $\tau^*\sim N^4$,
beyond which the data flatten out as they should. \label{fig3}}
\end{figure}

Since the modes are independent of each other at all times, using
Eqs. (\ref{e34a}-\ref{e34c}) we have
\begin{eqnarray} \langle\Delta \tilde{r}_m^2(\tau)\rangle=2 \sum_p
\frac{T^*}{4\zeta_p}
\left[\phi_p^{(N+1)/2}+\phi_p^{(N-1)/2}\right]^2\left[1-\exp(-\zeta_p
\tau)\right],
\label{e36a}
\end{eqnarray} which is evaluated in the Appendix. Only the even modes
contribute to the sum, and it is once again dominated by the
transverse modes, with the result that at intermediate times
\begin{eqnarray} \langle\Delta
\tilde{r}_m^2(\tau)\rangle=\Gamma(1/4)\frac{4 T^* \tau^{3/4}}{3 \pi
\nu^{1/4}},
\label{e37a}
\end{eqnarray} i.e., it increases subdiffusively in time with an
exponent $3/4$.  The subdiffusive behavior is seen until time
$\tau^*$, beyond which it saturates. The verification of this result
can be found in Fig. \ref{fig3}, which, like Fig. \ref{fig2}, also
demonstrates that the conversion of the mode sum (\ref{e31}) to
integrals suffers from strong finite-size effects; with increasing $N$
the mode sum (\ref{e31}) does approach the expected subdiffusive
behavior (\ref{e36a}).

The subdiffusive behavior with exponent $3/4$ can also be argued in
the following manner. In time $\tau_0\sim1$ the center-of-mass of the
two middlemost beads moves by a distance of $N^0$ due to the internal
motion, i.e. the MSD is of order $N^0$. Around $\tau^*\sim N^4$ the
center-of-mass motion starts to dominate and the MSD becomes order $D
\tau^* \sim \tau^*/N \sim N^3$. If the two values are to be bridged by
a single power-law, then the only possible exponent is $3/4$.

\subsection{Mode sums vs stretched exponents\label{sec4d}}
\begin{figure}[h]
\begin{minipage}{0.47\linewidth}
\includegraphics[width=\linewidth]{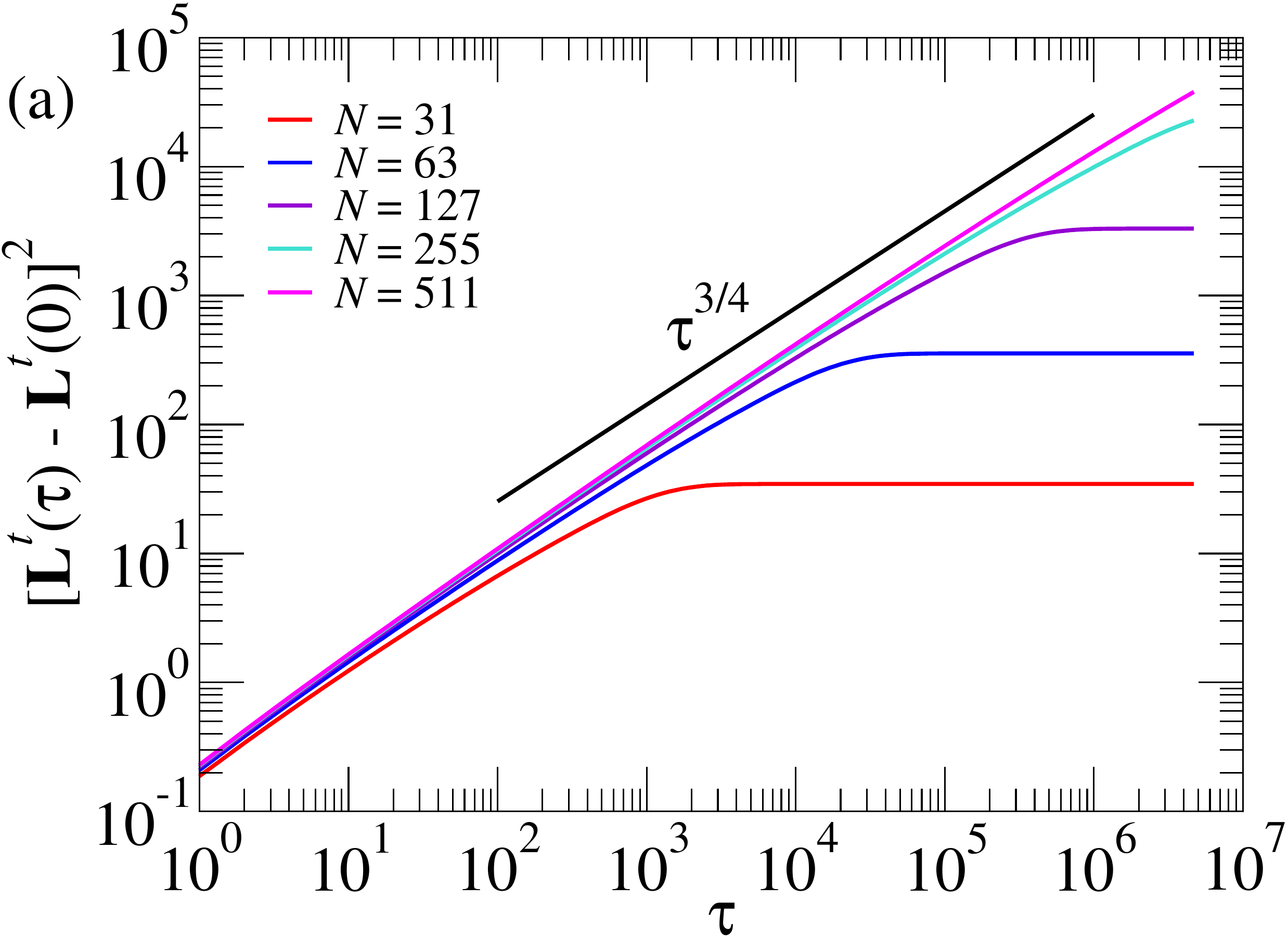}
\end{minipage} \hspace{5mm}
\begin{minipage}{0.47\linewidth}
\includegraphics[width=\linewidth]{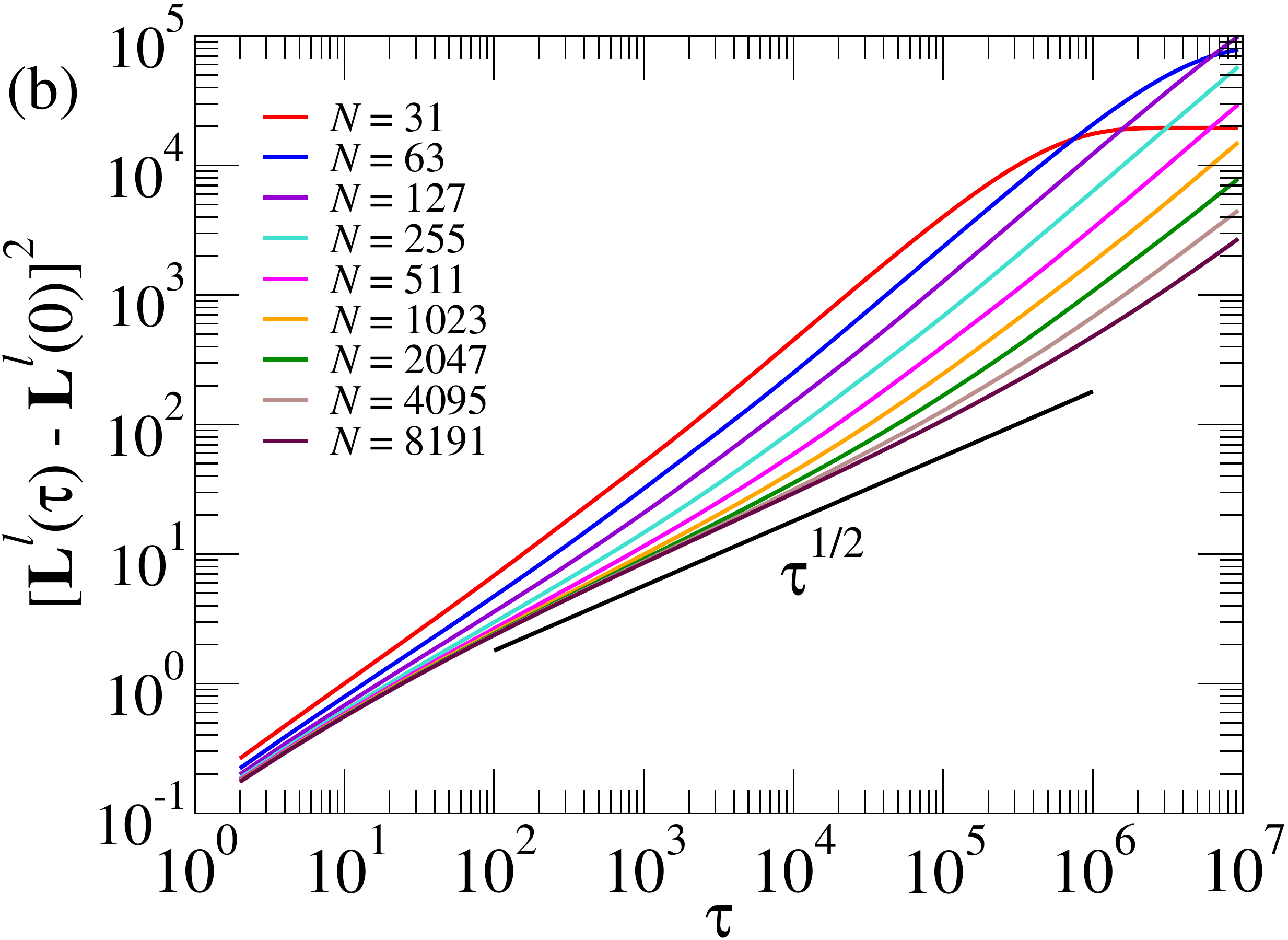}
\end{minipage}
\caption{(a) Log-log plot of the transverse MSD of the end-to-end
vector as given by the expression (\ref{e31}) for a series of lengths.
For comparison the anticipated power $\tau^{3/4}$ has been drawn.  (b)
Log-log plot of the longitudinal MSD of the end-to-end vector as given
by the expression (\ref{e31}) for a series of lengths. The convergence
to the anticipated power $\tau^{1/2}$ is too slow to be
useful.\label{figtranslong}}
\end{figure} In the previous subsections, we have extracted stretched
exponents from the mode sums by replacing the sums over the mode index
$p$ by integrals and approximating the modes by their low-$p$ behavior
as given in (\ref{m2}) and (\ref{m3}). Since our interest is in chain
lengths shorter than or comparable to the persistence length, it is
important to know how well the mode sums converge to this asymptotic
behavior.  In Fig.~\ref{figtranslong} we have plotted the transverse
mode sums for the MSD of the end-to-end vector, for a set of chain
lengths shorter and longer than the persistence length $l_p=114$ (for
the dsDNA parameters $T^*=0.034$ and $\nu=0.35$). The figure shows
that the mode sums, in the intermediate time regime $ \tau \ll
\tau^*$, are very well represented by the power $\tau^{3/4}$ over a
large time domain, also for chains of the order of the persistence
length.  The time domain, for which the stretched exponent holds,
expands with $N$.

In the same Fig.~\ref{figtranslong} the longitudinal mode sums for the
MSD of the end-to-end vector are plotted.  Here one observes that the
convergence to the anticipated power $\tau^{1/2}$ is very
slow. ($\tau^{1/2}$ follows from the replacement of the sum by an
integral). Even a chain of length of $N=8191$, which is about 75 times
the persistence length, cannot be represented in a substantial time
domain by the power $\tau^{1/2}$.  This is one of the reason that we
have concentrated in the previous sections on the analysis of the
transverse mode sums. The other reason is that the contribution of the
longitudinal modes in the total end-to-end vector remains small, since
the transverse modes dominate for intermediate times and the
orientational diffusion takes over for long times.

\subsection{Summary statements on linearized dynamics\label{sec5a}}

We now close off linearized semiflexible polymer dynamics with a
summary.

Based on the linearized dynamics for our model we have obtained
analytical expressions for (i) the autocorrelation function of the
end-to-end vector, (ii) the autocorrelation function of a bond (i.e.,
a spring, or a tangent) vector at the middle of the chain and (iii)
the mean-square displacement of a tagged bead in the middle of the
chain, as sum over the contributions from the transverse and/or
longitudinal modes --- the so-called mode sums.  The mode sum exhibit
the following asymptotic behavior.  (i) The end-to-end vector
autocorrelation function for the chain decays in time as a stretched
exponential with an exponent $3/4$, crossing over to pure exponential
decay at the terminal time $\tau^*=(\zeta_2^t)^{-1}$.  (ii) The
autocorrelation of the orientation of the middlemost bond vector
decays in time in a similar manner, but with an exponent $1/4$.  (iii)
The mean-square displacement (MSD) of the middle bead shows anomalous
diffusion with an exponent $3/4$ until time $\tau^*$, beyond which its
motion becomes diffusive.  The convergence with increasing $N$ to the
asymptotic power is fast for the transverse mode sum and very slow for
the longitudinal mode sum.

Further, dynamical quantities as obtained from mode sums show a
remarkable agreement with numerical simulation results of dsDNA chains
with lengths shorter or of the order of the persistence length. The
main exceptions to this agreement stem from the fact that if the chain
bends, it tends to conserve its curvilinear length, thereby reducing
the distance between the ends. The mode sums do not capture this
effect, which is nonlinear and a consequence of coupling between the
transverse and longitudinal modes. This is discussed in more detail in
Sec. \ref{sec5}.

\section{Separating the longitudinal and transverse
components} \label{sec5}

In the previous section we have compared the results of linearized
dynamics with the simulations, involving the full polymer dynamics of
the model in terms of three correlation functions. The observed
excellent agreement suggests that for lengths less than $l_p$ the
linearized dynamics suffices, which opens up a wide avenue of
opportunity for analytical calculations using our model. In this and
the next sections we will examine this agreement in further detail,
taking the end-to-end vector as an example.

The end-to-end vector correlation function calculations in
Fig.~\ref{fig1} (and other quantities) did not need a distinction
between longitudinal and transverse fluctuations, hence the
simulations for these quantities can be carried out with any
integration scheme, including the integration of equations
(\ref{hesseq}) employing the positions of the beads. As we now embark
on separating the longitudinal and transverse fluctuations, the
polymer dynamics in the simulations requires more care, in particular
the role of the coupling forces and the orientation of the
groundstate, as will be described below.

The coupling force arises from the derivative of the contour length
$L_c$ and in terms of the mode representation, we have
\begin{eqnarray} {\bf H}_p = \sum_n \left[ \frac{\partial
L_c}{\partial {\bf r}_n} - \left(\frac{\partial L_c}{\partial {\bf
r}_n}\right)^{(0)} \right] \, \phi_{n,p}.
\label{e50}
\end{eqnarray} Having worked out the partial derivatives we find for
the longitudinal component
\begin{equation} {\mathbf H}^l_p = \sum_n \left(\hat{\bf u}_n -
\hat{\bf u}^{(0)}_n \right)^l \left[\phi^l_{n,p} -
\phi^l_{n-1,p}\right]
\label{e51}
\end{equation} and for the transverse components
\begin{equation} {\bf H}^t_p =\sum_n {\bf u}^t_n \left(\frac{1}{u_n}-
\frac{1}{u^{(0)}_n } \right) \left[\phi^t_{n,p} - \phi^t_{n-1,p}
\right].
\label{e52}
\end{equation} It is worthwhile to note that the longitudinal coupling
force arises from the transverse fluctuations --- a pure longitudinal
deformation will not change the direction $\hat{\bf u}_n$, as
occurring in Eq. (\ref{e51}), from that in the groundstate $\hat{\bf
u}^{(0)}_n$.  This is also reflected in the fact that the longitudinal
modes are exact eigenmodes of the system in the groundstate.

The simulation is carried out by forward-integrating Eq. (\ref{e10}).
For the integration we represent the chain configurations by
continuously alternating between the mode representation (for the
timestep) and the position representation (for the calculation of the
coupling forces).  For the longitudinal components the transformation
back and forth between the mode representation and position
representation is a (fast) Fourier Transform. For the transverse
components the back and forth transformations are found using the
transverse eigenfunctions $\phi^t_{n,p}$.

As discussed in Sec. \ref{sec2c}, the groundstate is identified by
enforcing the components of the transverse modes for $p=1$ strictly
equal to zero. This condition must be satisfied, in principle, at
every time step, requiring an adjustment of the orientation of the
groundstate, by rotating the direction $\hat{\bf e}_0$ around an axis
$\bf \omega$ perpendicular $\hat{\bf e}_0$. The adjustment of the
orientation of the groundstate leaves the spatial configuration of the
beads invariant, but induces a transformation of its representation in
modes. The direction of $\bf \omega$ as well as details of the
transformation are provided in Appendix \ref{adjust}.
\begin{figure}[h]
\begin{minipage}{0.48\linewidth}
\includegraphics[width=\linewidth]{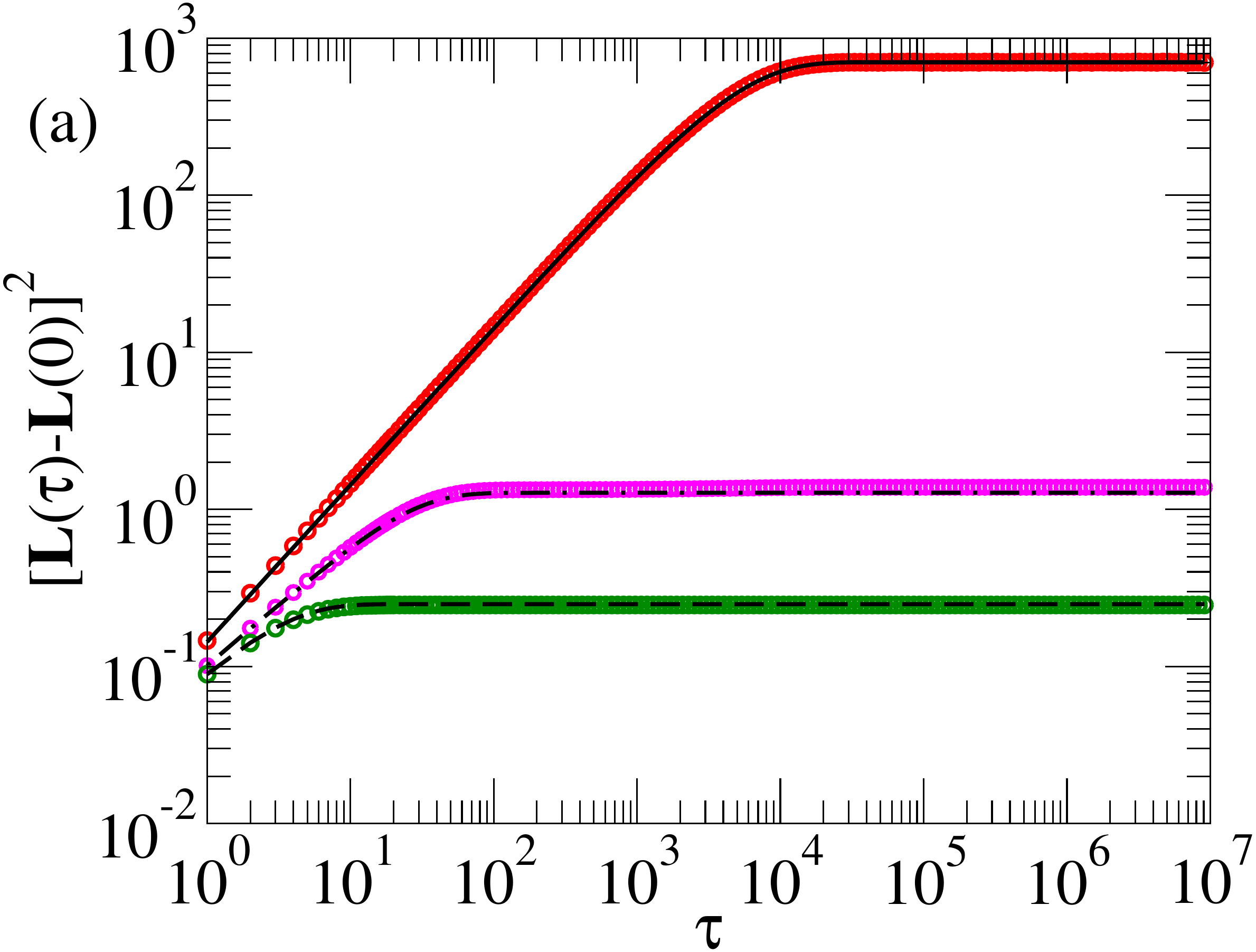}
\end{minipage} \hspace{5mm}
\begin{minipage}{0.46\linewidth}
\includegraphics[width=\linewidth]{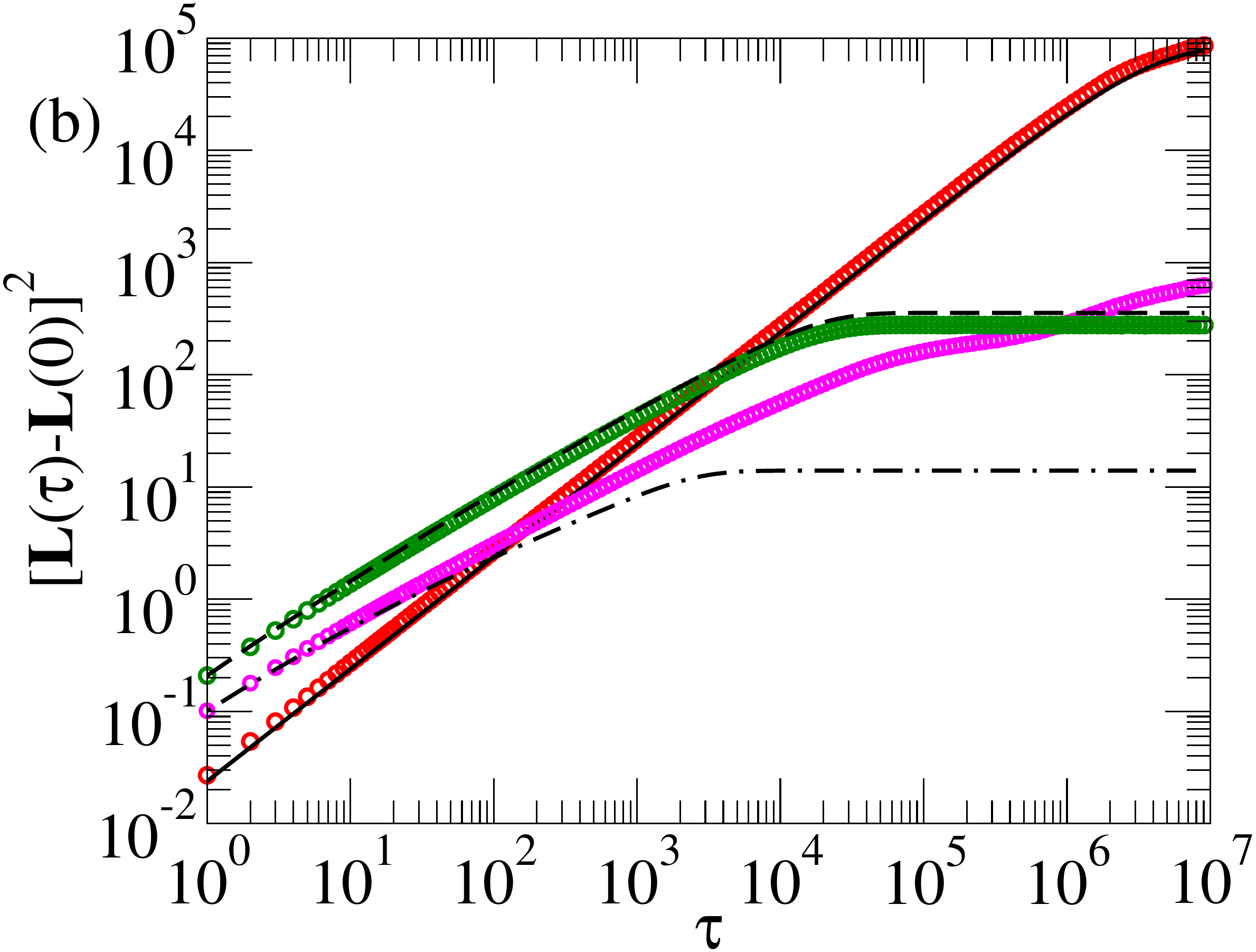}
\end{minipage}
\caption{The orientational, longitudinal and transverse components of
the mean-square displacements $[{\bf L}^{(0)}(\tau)-{\bf
L}^{(0)}(0)]^2$ (linearized theoretical results in solid lines and
simulation data in red symbols), $[{\bf L}^l(\tau)-{\bf L}^l(0)]^2$
(linearized theoretical results in dash-dotted lines and simulation
data in magenta symbols), and $[{\bf L}^t(\tau)-{\bf L}^t(0)]^2$
(linearized theoretical results in dashed lines and simulation data in
green symbols) respectively: (a) $N=7$, (b) $N=63$. \label{fig4}}
\end{figure}

The linearized theoretical results and the simulation data for the
orientational, longitudinal and transverse components of the
mean-square displacements $[{\bf L}^{(0)}(\tau)-{\bf L}^{(0)}(0)]^2$,
$[{\bf L}^l(\tau)-{\bf L}^l(0)]^2$ and $[{\bf L}^t(\tau)-{\bf
L}^t(0)]^2$ respectively [see Eq. (\ref{e19}) for the definitions] are
compared for $N=7$ and $63$ for dsDNA in Fig.~\ref{fig4}. For very
short chains (such as $N=7$) the coupling force stays small in
amplitude, explaining the excellent agreement between the simulations
and the linearized dynamics results. For $N=63$ however, we see that
at long times, the agreement between the linearized theory and the
simulation is rather poor for the longitudinal component $[{\bf
L}^l(\tau)-{\bf L}^l(0)]^2$.  Despite this disagreement, we clearly
see in Fig.~\ref{fig4}(b) that in the region of the strongest
disagreement the orientational component $[{\bf L}^{(0)}(\tau)-{\bf
L}^{(0)}(0)]^2$ --- for which the linearized theoretical results and
the simulation data {\it do\/} agree very well --- is two orders of
magnitude stronger than the longitudinal component $[{\bf
L}^l(\tau)-{\bf L}^l(0)]^2$. This observation is therefore consistent
with the good agreement between the linearized theoretical results and
the simulation data for $C_L(\tau)$, wherein all the orientational,
longitudinal and transverse components combine together.

Figure \ref{fig4}(b) shows that the main deviations from the linear
theory are in the longitudinal component of the end-to-end vector for
asymptotic large times. This is a result of the coupling between the
longitudinal and the transverse modes. The bending of the chain due to
transverse fluctuations shortens the end-to-end distance in the
longitudinal direction.  The next section is devoted to a further
analysis of these effects.

\section{Non-linear effects in semiflexible polymer
dynamics\label{sec6}}

Semiflexible polymer dynamics is inherently nonlinear. If the
nonlinear effects get strong then they clearly ruin the agreement
between the (linearized) mode sums and the simulation data. An
interesting question is, why do they show up strongly in longitudinal
fluctuations, e.g., in the quantity $[{\bf L}^l(\tau)-{\bf
L}^l(0)]^2$?

There are two different classes of situations where we expect the
nonlinearities to become strong. (i) The first case is when the chain
gets long in comparison to its persistence length. In this case the
transverse fluctuations become progressively easier to excite with
increasing chain lengths, and the effective chain length along the
vector $\hat{\mathbf e}_0$ that denotes the orientation of the
groundstate shortens from ${\bf L}^{(0)}_0$.  (ii) The second case is
when the chain gets progressively more inextensible. In the limit when
the chain is completely inextensible, like the WLC model, any
transverse fluctuation results in an immediate shortening of the
end-to-end distance. This case is best analyzed by reducing $T^*$,
i.e., reducing the stretchability of the bonds, while simultaneously
keeping the persistence length $l_p=\nu/T^*$ fixed. In this case one
expects strong nonlinearities to emerge also in chains that are short
with respect to the persistence length. Both cases affect longitudinal
fluctuations the most, hence the quantity $[{\bf L}^l(\tau)-{\bf
L}^l(0)]^2$ is the most sensitive to nonlinearities in the model.

\begin{figure}[t]
\begin{minipage}{0.48\linewidth}
\includegraphics[width=\linewidth]{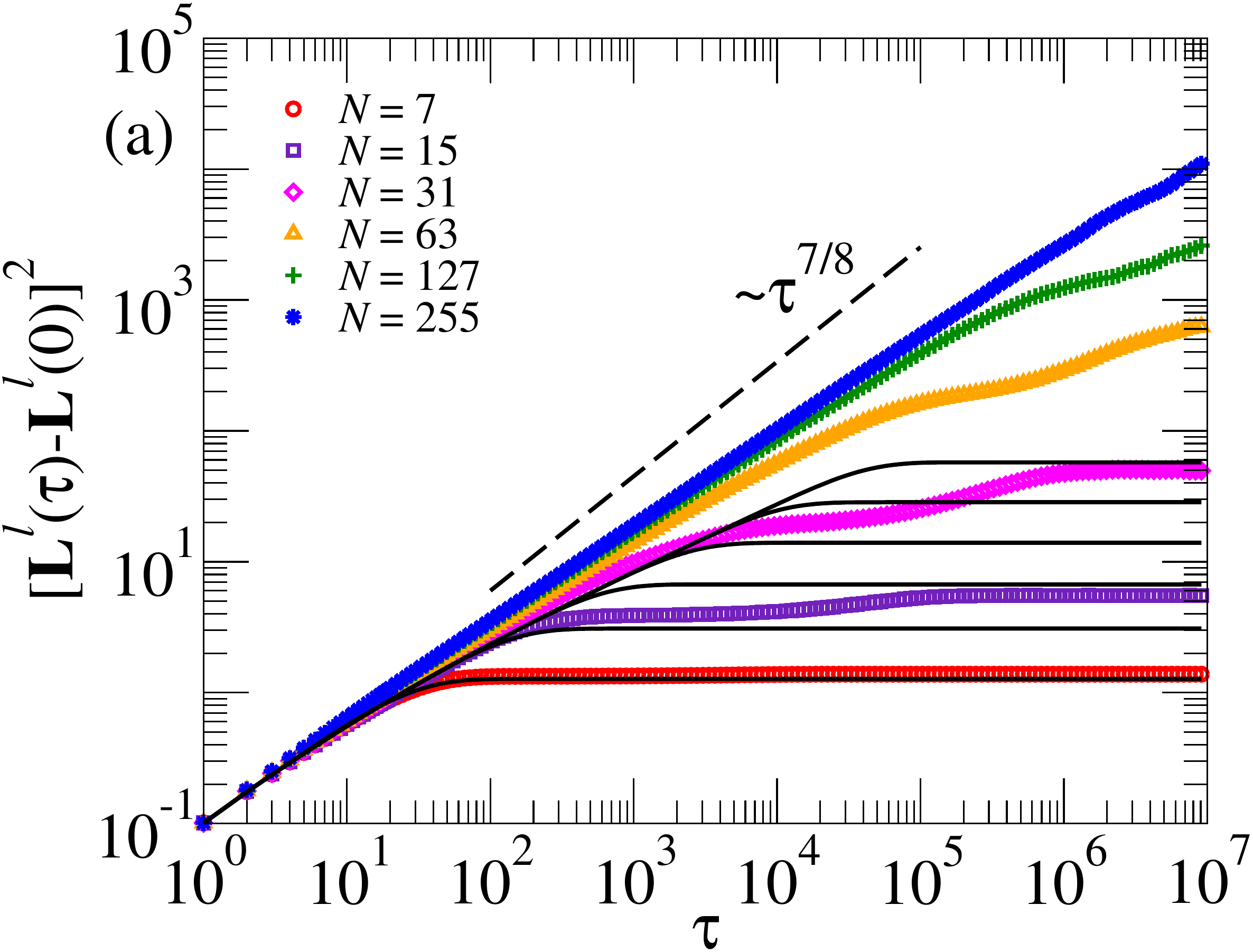}
\end{minipage} \hspace{5mm}
\begin{minipage}{0.46\linewidth}
\includegraphics[width=\linewidth]{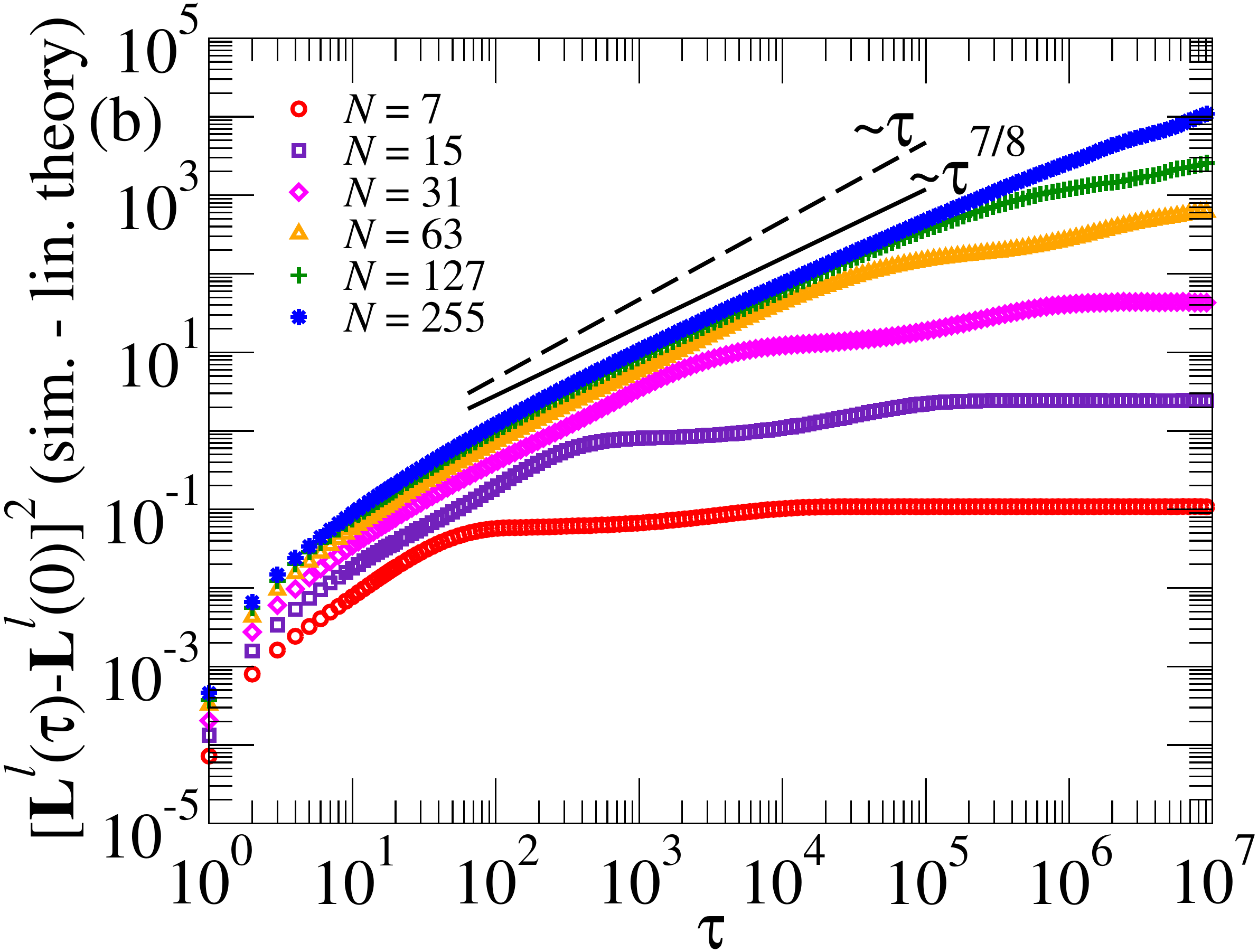}
\end{minipage}
\caption{Plots for the longitudinal component of $[{\bf
    L}^l(\tau)-{\bf L}^l(0)]^2$ for dsDNA for $N=7,15,31,63,127$.  and
  $255$. (a) The total value of $[{\bf L}^l(\tau)-{\bf L}^l(0)]^2$ is
  plotted; symbols denote simulation data, and solid lines denote the
  corresponding linearized mode sums for the same values of $N$ as in
  the simulations (the value of $N$ corresponding to the mode sums
  increase from bottom to top). The dashed line shows corresponds to
  the power-law $\tau^{7/8}$. (b) The pure non-linear effect as the
  difference between the simulated values and the linear contribution
  given by the mode sum. In the difference the emergence of an
  effective power-law $\tau^{7/8}$ (solid line) can be seen at
  intermediate times before leveling off. For comparison, also plotted
  is a dashed line showing power-law $\tau^{1}$.\label{glongdna}}
\end{figure}

We first discuss case (i) as a follow-up of $[{\bf L}^l(\tau)-{\bf
L}^l(0)]^2$ in Fig.~\ref{fig4} for dsDNA. In order to clearly see the
non-linear effects we have plotted in Fig.~\ref{glongdna} the simulated 
values as well as the {\it
difference} between the simulated values and that given by the
corresponding linearized dynamics theory. We see that for short chains
the difference remains small, while it grows to substantial values for
long chains. This is because the linearized theory predicts that
$[{\bf L}^l(\tau)-{\bf L}^l(0)]^2$-values reach a plateau beyond the
lifetime of the longitudinal modes as seen in Fig.~\ref{fig4}, but the
simulated values keep increasing due to nonlinearities [encoded in the
coupling force term as described in Eq.~(\ref{e50})]. As already
explained above, the main source of this non-linear effect is the
shortening of the end-to-end distance along the vector $\hat{\mathbf
e}_0$ due to the transverse fluctuations. Indeed, in support of the
arguments given in the above paragraph, we find that with increasing
chain lengths the longitudinal modes, in particular those with a low
odd index $p$, do not fluctuate around zero any more (as assumed in
the linearized dynamics), but around a positive average. If we
translate these positive averages back to spatial positions, we find a
shorter distance $\langle ({\bf r}_N - {\bf r}_0) \cdot \hat{\bf e}_0
\rangle$ than the groundstate value $L^{(0)}$. The shortening of the
distance becomes for longer chains much larger than the amplitude of
the longitudinal fluctuations.

An interesting feature of the curves of Fig.~\ref{glongdna} is that
$[{\bf L}^l(\tau)-{\bf L}^l(0)]^2$ at intermediate times is a
combination of the (linear) mode sums and inherently nonlinear
behavior of the model. Indeed, the pure nonlinear contributions to
$[{\bf L}^l(\tau)-{\bf L}^l(0)]^2$ to an effective power-law
$\tau^{7/8}$ at intermediate times before the data level off, as
indicated by the straight line in the log-log plot in
Fig. \ref{glongdna}(b). The mode sum contributions, on the other hand,
although very slowly converges to $\tau^{1/2}$ behavior as seen in
Fig. \ref{figtranslong}, we expect an effective exponent less than
$7/8$ for $[{\bf L}^l(\tau)-{\bf L}^l(0)]^2$, as confirmed in
Fig. \ref{glongdna}(a).  (We mention in passing here that the
nonlinear effects in the longitudinal fluctuations are well-documented
in the WLC literature. E.g., for (inextensible) WLC model, a power law
in time with exponent $7/8$ is reported for the fluctuations in the
longitudinal component of the end-to-end vector
\cite{hall2,liverpool4,ober2} --- this prompts the comparison of our
data in Fig.~\ref{glongdna} to a power-law $\tau^{7/8}$. We will
return to this in the next section.)
\begin{figure}[h]
\begin{minipage}{0.48\linewidth}
\includegraphics[width=\linewidth]{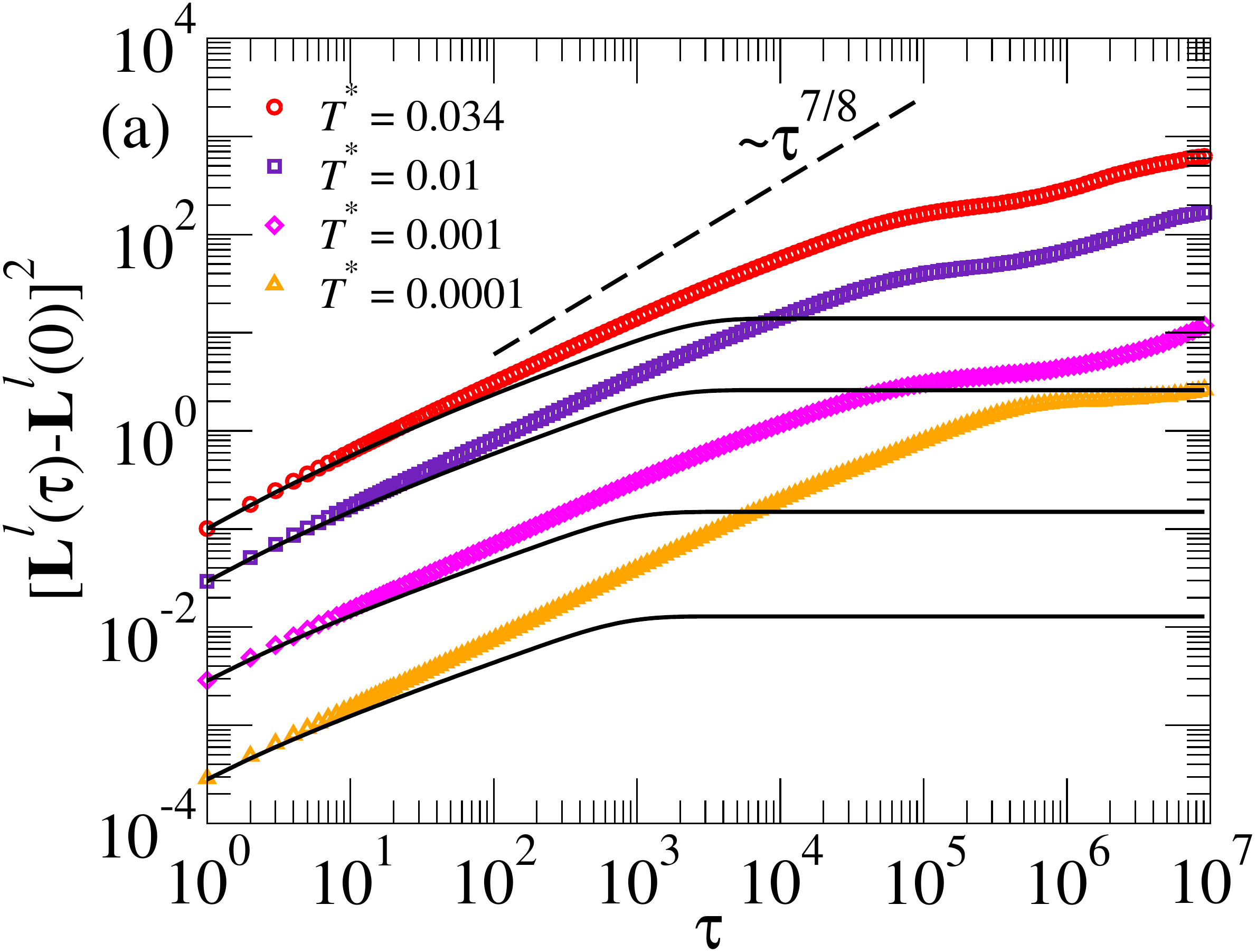}
\end{minipage} \hspace{5mm}
\begin{minipage}{0.46\linewidth}
\includegraphics[width=\linewidth]{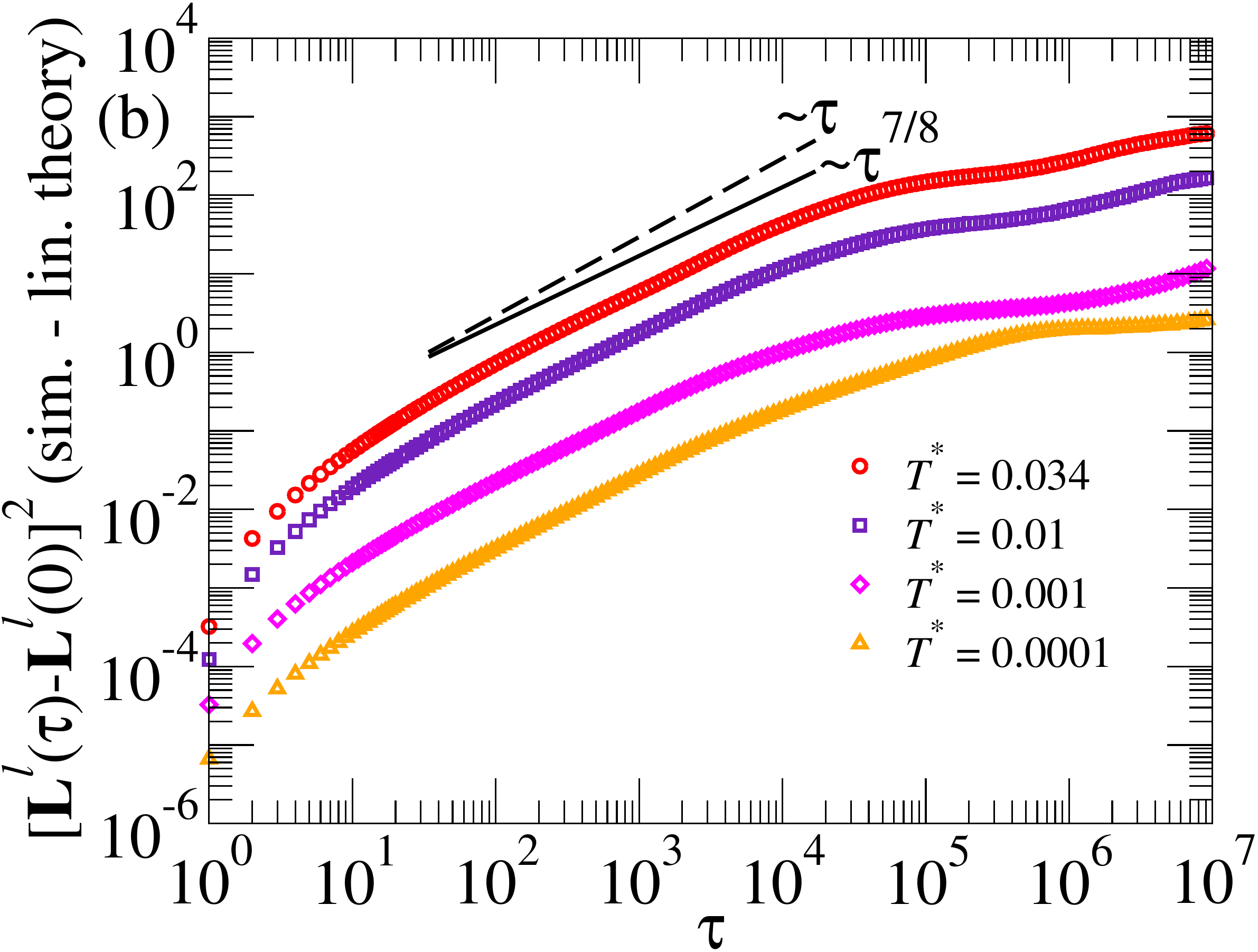}
\end{minipage}
\caption {The longitudinal component of the end-to-end vector $[{\bf
    L}^l(\tau)-{\bf L}^l(0)]^2$ for dsDNA ($T^*=0.034$) and a few
  other values of $T^*$. Here $N=63$ and the persistence length is
  $\approx114$ nm for all cases. As in Fig. \ref{glongdna} we plot in
  part (a) the total value of $[{\bf L}^l(\tau)-{\bf L}^l(0)]^2$;
  symbols denote simulation data, and solid lines denote the
  corresponding linearized mode sums for the same values of $T^*$ as in
  the simulations (the value of $T^*$ corresponding to the mode sums
  increase from bottom to top), and the dashed line shows corresponds
  to the power-law $\tau^{7/8}$. in part (b) the difference between
  the simulation and the linear contribution as given by the mode
  sum. Again in the difference the emergence of an effective power-law
  $\tau^{7/8}$ (solid line) can be seen at intermediate times before
  leveling off. For comparison, also plotted is a dashed line showing
  power-law $\tau^{1}$.\label{hasplit}}
\end{figure}

Case (ii) yields a similar picture. In the limit of small $T^*$ (and
$l_p=\nu/T^*$ still fixed at its dsDNA value $\approx114$ nm), the
contribution of the longitudinal fluctuations within the linearized
theory vanishes with the power $T^*$, since the decay of the modes
becomes independent of $\nu$ for $\nu \rightarrow 0$. However the
transverse modes, for which the decay coefficients are proportional to
$\nu$ [c.f. Eq. (\ref{m3})], survive longer being proportional to
$T^*/\nu$. Since the transverse modes for more unstretchable chains
imply a shortening of the end-to-end distance with respect to the
groundstate, the longitudinal modes again get a non-vanishing average
in equilibrium. This is shown in Fig. \ref{hasplit}, where we have
plotted the longitudinal component of the end-to-end vector for chain
length $N=63$ with $l_p\approx114$ nm not only for dsDNA
($T^*=0.034$), but also for $T^*=0.01, 0.001$ and $0.0001$. Again we
plot $[{\bf L}^l(\tau)-{\bf L}^l(0)]^2$ as well as the difference
between the simulation and the linear contribution as given by the
mode sum, and the total. The $[{\bf L}^l(\tau)-{\bf L}^l(0)]^2$ data
show an effective exponent less than $7/8$, while the pure nonlinear
contributions to $[{\bf L}^l(\tau)-{\bf L}^l(0)]^2$, although
initially very small with respect to the total contribution, they grow
with time, as an effective power-law $\tau^{7/8}$ at intermediate
times, before saturating to different plateau-values for different
values of $T^*$. We note that it difficult to extract from the total
contribution an effective power law, since the linear contribution
suffers from large finite size effects as Fig. \ref{figtranslong}
shows.

Our results for $[{\bf L}^l(\tau)-{\bf L}^l(0)]^2$ bear strong
resemblance to those of Refs. \cite{recent6,ober2} where the
extensible WLC has also been shown to demonstrate the exponents $1/2$
and $7/8$ at intermediate times in the dynamic elastic modulus,
obtained from the (length fluctuation) response of the chain to a
(fluctuating) tensile force. Linear response theory suggests that the
dynamic elastic modulus via a memory kernel, must relate to the
equilibrium fluctuations in the longitudinal end-to-end vector studied
by ourselves. For tagged monomer displacement, such a case has been
studied in detail by one of us \cite{panja1,panja2} by means of
Generalized Langevin Equation (GLE) formulation. In the present case
however, we do not know how to quantitatively relate our results for
the mean-square displace longitudinal end-to-end vector and those of
Refs. \cite{recent6,ober2} using a GLE, or some other, formulation.

We again emphasize here that even at the times when the nonlinear
contribution to $[{\bf L}^l(\tau)-{\bf L}^l(0)]^2$ become significant,
it still remains typically two orders of magnitude smaller than the
orientational component $[{\bf L}^{(0)}(\tau)-{\bf L}^{(0)}(0)]^2$, so
nonlinearities make little difference for the quantities dependent on
the overall end-to-end vector. It is in this sense we state, in view
of Figs. \ref{fig1}-\ref{fig4} that the linearized dynamics suffices
very well for dsDNA fragments that are shorter than or comparable to
the persistence length.

\section{Conclusion\label{sec7}}

With a recently introduced Hamiltonian bead-spring model for
semiflexible polymers \cite{hans}, in this paper we have studied the
dynamical properties of a chain of length $N$. Specifically, using
linearized polymer dynamics we have analytically calculated the
autocorrelation functions of the end-to-end vector, that of the
orientation of the middle bond and the mean-square displacement of the
middlemost bead of the chain. The analytical solutions are facilitated
by the fact that we know the dynamical mode structures analytically in
the Hessian approximation of the Hamiltonian. There are longitudinal
and transverse modes, and the terminal time $\tau^*\sim N^4$ is
determined as the inverse of the smallest eigenvalue $\zeta^t_2$ for
the transverse modes. Up to time $\tau^*$ the autocorrelation
functions for the end-to-end vector and that of the middle spring
vector show stretched exponential behavior with exponents $3/4$ and
$1/4$ respectively, and beyond time $\tau^*$ the decay becomes simply
exponential. The motion of the middlemost bead exhibit anomalous
dynamics with an exponent $3/4$ until time $\tau^*$, and is diffusive
thereafter.

The full dynamics of semiflexible polymers in our model is obviously
nonlinear, and we have also performed simulations of the full dynamics
for chains. We find that for dsDNA the mode sums agree remarkably well
with the numerical values obtained from simulations. This does not
however mean that the nonlinearities are not present. Indeed, we find
that the MSD of the longitudinal component of the end-to-end vector
showcases strong nonlinear effects in the polymer dynamics, and we
identify at least an effective $t^{7/8}$ power-law regime in its
time-dependence. We show that the nonlinear effects in the MSD of the
longitudinal component of the end-to-end vector increases with
increasing length for dsDNA; nevertheless, in comparison to the full
mean-square displacement of the end-to-end vector the nonlinear
effects remain small at all times. It is in this sense we state that
the linearized dynamics suffices for dsDNA fragments that are shorter
than or comparable to the persistence length.

Given that the worm-like chain (WLC) is the most used model for
semiflexible polymers, it is imperative to compare the results of the
above quantities across the two models. First of all, the anomalous
dynamics of the middlemost bead agree for the two models
\cite{kroy,farge}, and have also been found in experiments
\cite{expt1,expt2,expt3}. On the other hand, the end-to-end vector
correlation function has been considered in Ref. \cite{liverpool4},
wherein $\langle[{\mathbf L}(t)-{\mathbf L}(0)]^2\rangle$ is seen to
behave as $A_1t^{3/4}+A_2t^{7/8}$, with $A_1$ and $A_2$ two
$N$-dependent constants, of which the $t^{7/8}$ term is known to
originate from nonlinear effects in WLC dynamics, and showcases itself
in the longitudinal fluctuations of the end-to-end vector. The form of
the term proportional to $A_1$ is consistent with the stretched
exponential behavior (\ref{e37}) when $\tau\ll\tau^*$ [since
$\exp(-Bx^\alpha)\approx1-Bx^\alpha$ for $Bx^\alpha\ll1$]. Further, we
have also found at least an apparent $t^{7/8}$ power-law in the
longitudinal MSD of the end-to-end vector, albeit with an almost
insignificant amplitude wrt the overall fluctuations of the end-to-end
vector (we cannot ascertain a true power-law behavior since we do not
have an analytical derivation for $t^{7/8}$). Further, Ref.
\cite{liverpool4} reports the orientational correlation function of
the unit tangent vector at the middle of the chain, which is analogous
to that of the orientation of the middle bond vector ${\mathbf u}_m$
in our model. The corresponding result, translated in terms of
${\mathbf u}_m$ would imply that $\langle[{\mathbf u}_m(t)-{\mathbf
  u}_m(0)]^2\rangle=B_1t^{1/4}+B_2t^{1/2}$, with $B_1$ and $B_2$ two
$N$-dependent constants. The form of the term proportional to $B_1$ is
consistent with the stretched exponential behavior (\ref{e37}) at
$t\ll\tau^*$, but we do not find any signature of an exponent $1/2$
for $\langle[{\mathbf u}_m(t)-{\mathbf u}_m(0)]^2\rangle$ in our
result. It is of course possible that the $t^{1/2}$ behavior arises
from the (nonlinear) longitudinal fluctuations, which, in light of the
longitudinal fluctuations in the end-to-end vectors, we also expect to
have an almost insignificant impact on the total fluctuations in
${\mathbf u}_m(t)$. We have already shown results for $N=255$ for
dsDNA in one occasion (\ref{glongdna}) --- an instance where we are
able to simulate, at the basepair resolution, full chains of length
more than twice the persistence length.

\appendix
\section{The stretched exponentials derived from the
mode-sums}\label{stret}

\setcounter{equation}{0}
\renewcommand{\theequation}{A\arabic{equation}}

In this appendix we present some formulas that enable us to analyze
$C_L(t)$. We start with sums of the type
\begin{eqnarray} F_n (\tau) = \sum_{p=3,} \frac{p^n}{\zeta^t_p} [1-
\exp (-\zeta^t_p \tau)]),
\label{A1}
\end{eqnarray} The function $F_0 (\tau)$ is relevant for the
auto-correlation function of the end-to-end vector and the mean
squared displacement of the middle bead.  The function $F_2 (\tau)$ is
needed for the auto-correlation function of the middle bond, and for
next to dominant contributions.  The first step is to replace the sum
by an integral. The justification comes from the fact that for small
values of the exponent (which is the case for $\tau \leq \tau^*= N^4$)
a large range of $p$-values contribute, for which the integrand is
smoothly varying. So, we approximate $F_0 (\tau)$ by
\begin{eqnarray} F_n (\tau) \simeq \frac{1}{2} \int dp
\frac{p^n}{\zeta^t_p} [1- \exp (-\zeta^t_p \tau)]),
\label{A2}
\end{eqnarray} Next we approximate $\zeta^t_p$ in the regime where the
main contributions come from by
\begin{eqnarray} \zeta^t_p \simeq \frac{\nu \pi^4 p^4}{N^4}
\label{A3}
\end{eqnarray} and make the substitution
\begin{eqnarray} p = \frac{N}{\pi (\nu \tau)^{1/4}} q^{1/4},
\label{A4}
\end{eqnarray} which leads for $F_0(\tau)$ to the integral
\begin{eqnarray} F_0 (\tau) \simeq \frac{N \tau^{3/4}}{8 \pi
\nu^{1/4}} \int dq \, q^{-7/4} [1- \exp (-q)].
\label{A5a}
\end{eqnarray} Integration by parts gives the result
\begin{eqnarray} F_0 (\tau) \simeq \frac{N \tau^{3/4}}{6 \pi
\nu^{1/4}} \int dq \, q^{-3/4} \exp (-q)= \frac{N \tau^{3/4}}{6 \pi
\nu^{1/4}} \Gamma(1/4).
\label{A5}
\end{eqnarray}

For $F_2(\tau)$ the derivation is similar. Only the front factor and
the power of $\tau$ differ:
\begin{eqnarray} F_2 (\tau) \simeq \frac{N^3 \tau^{1/4} }{2 \pi^3
\nu^{3/4}} \Gamma(3/4).
\label{A6}
\end{eqnarray}

In Section \ref{sec4a} the interplay between the front factors and the
time dependence has been discussed with the result that times $\tau
\sim N^4$ are most relevant for the exponent of the stretched
exponential. As follows, one can see that the next order $\sim p^2$ in
the expansion of $L_p$ is dwarfed by Eq. (\ref{A5}). The term has an
extra factor $N^2$ in the denominator and gets another extra factor
$N^2$ in the numerator as $F_0$ has to be replaced by $F_2 $ (with the
second extra factor). But $\tau^{3/4}$ in $F_0$ gives a factor $N^3$
for $\tau \sim \tau^*$ and $\tau^{1/4}$ in $F_2$ gives a factor $N$.
So the next term in the expansion is a factor $N^{-2}$ smaller than
the dominant term.

The contribution of the longitudinal modes cannot be observed for
similar reasons. The analysis with the spectrum $\zeta^l_p \sim
(p/N)^2$ gives the combination $\tau^{1/2} N^{-2}$, which applies for
times bounded by $(\zeta^l_1) \sim N^2$. Hence, the combination
$\tau^{1/2} N^{-2}$ remains very small in that time regime.
 
\section{Adjustment of the reference groundstate\label{adjust}}

\setcounter{equation}{0}
\renewcommand{\theequation}{B\arabic{equation}}

In this appendix we discuss the adjustment of the reference
groundstate such that the transverse mode $p=1$ are kept equal to
zero. The adjustment amounts to a rotation of the vectors ${\bf
r}^0_n$ over an angle $\theta$ around an axis $\hat{\bf \omega}$. With
this $\hat{\bf \omega}$ and $\theta$ the set of reference axes
$\hat{\bf e}_\alpha$ are rotated to a system $\hat{\bf e}^R_\alpha$
connected to the original ones by the matrix
\begin{equation} \label{B1} \hat{\bf e}^R_\alpha = \sum_\beta
T_{\alpha,\beta} \, \hat{\bf e}_\beta.
\end{equation} The matrix $T_{\alpha,\beta}$ is related to the
rotation $\hat{\bf \omega}, \theta$ by
\begin{equation} \label{B2} T_{\alpha,\beta} = \cos \theta \,
\delta_{\alpha,\beta} + (1-\cos \theta) \, \omega_\alpha \omega_\beta
+ \sin \theta \, \omega_{\alpha \times \beta}.
\end{equation} Here $\omega_\alpha$ is the component $\alpha$ of the
vector $\hat{\bf \omega}$.  The index $\alpha \times \beta$ means for
$\alpha \neq \beta$ the next one in the periodic series $0,1,2,0,1,
\dots$ with $\omega_{\beta \times \alpha} =- \omega_{\alpha \times
\beta}$. As $\hat{\omega}$ is a unit vector one has the relation
\begin{equation} \label{B3} \sum_\alpha \, \omega^2_\alpha = 1.
\end{equation} There is no point of rotating the vectors ${\bf r}^0_n$
around their common direction, so we put $\omega_0=0$.

The problem is to find the two other components $\omega_1$ and
$\omega_2$ which have the role to let the modes $\Psi^1_1$ and
$\Psi^2_1$ vanish. First we derive the transformation of the
components of the modes under a general rotation of the reference
basis.  Note that we rotate the reference basis but keep the
positions of the monomers fixed. As the longitudinal and transverse
mode behave differently we treat them separately. According to
(\ref{e10}) the longitudinal mode is related to the positions in the
rotated reference systems as
\begin{equation} \label{B4} (\Psi^0_p)^R = \sum_n \left[{\bf r}_n
\cdot \hat{\bf e}^R_0 - r^0_n\right] \phi^l_{n,p} = \sum_n
\left[\sum_\beta T_{0,\beta} \, ({\bf r}_n \cdot \hat{\bf e}_\beta) -
r^0_n \right] \phi^l_{n,p}.
\end{equation} We express, with (\ref{e9}), the components of
positions ${\bf r}_n$ back into the modes of the original reference
system
\begin{equation} \label{B5} {\bf r}_n \cdot \hat{\bf e}_\beta = r^0_n
\, \delta_{\beta,0} + \sum_q \phi^\beta_{n,q} \, \Psi^\beta_q.
\end{equation}

The longitudinal and transverse eigenfunctions are orthogonal for the
same type
\begin{equation} \label{B7} \sum_n \phi^l_{n,p} \, \phi^l_{n,q}
=\delta_{p,q} \quad \quad \sum_n \phi^t_{n,p} \, \phi^t_{n,q}
=\delta_{p,q},
\end{equation} but the mixed combination yields the matrix
\begin{equation} \label{B8} \sum_n \phi^l_{n,p} \, \phi^t_{n,q} =
A_{p,q},
\end{equation} which is nearly diagonal and which has only even-even
and odd-odd elements.  So we get for the longitudinal modes
\begin{equation} \label{B9} (\Psi^0_p)^R = \sqrt{I} (1-T_{0,0}) A_{p,1} +
T_{0,0} \, \Psi^0_p + \sum_{q, \beta=1,2} T_{0,\beta} \, A_{p,q}
\,\Psi^\beta_q.
\end{equation} We used for the summation over $n$ the explicit form of
the eigenfunction of the transverse mode $p=1$ given in (\ref{e8a}).

The transverse component transform according to (for $\alpha \neq 0$)
\begin{equation}
\label{B10} (\Psi^\alpha_p)^R = \sum_n \phi^t_{n,p} ({\bf r}_n \cdot
\hat{\bf e}^R_\alpha) = \sum_n \phi^t_{n,p} \sum_\beta
T_{\alpha,\beta} ({\bf r}_n \cdot \hat{\bf e}_\beta).
\end{equation} For the inner product we find
\begin{equation} \label{B11} {\bf r}_n \cdot \hat{\bf e}_\beta =
\delta_{\beta,0} (r^0_n + \sum_q \phi^l_{n,q} \Psi^0_q) +
(1-\delta_{\beta.0}) \sum_q \phi^t_{n,q} \Psi^\beta_q.
\end{equation} Inserting (\ref{B11}) into (\ref{B10}) yields, using
(\ref{e8a}) and (\ref{B7}),
\begin{equation} \label{B12} (\Psi^\alpha_p)^R = T_{\alpha,0}
\,\left(-\sqrt{I} \, \delta_{p,1} + \sum_q A_{p,q} \Psi^0_q \right) +
\sum_{\beta=1,2} T_{\alpha,\beta} \Psi^\beta_p.
\end{equation}

We get an equation for the components $\omega_1$ and $\omega_2$ by
requiring that $(\Psi^1_1)^R = (\Psi^2_1)^R = 0$. In order to make these
equations explicit we introduce the combinations
\begin{equation} \label{B13} u = \frac{\Psi^1_1}{\sqrt{I} - \sum_q
A_{p,q} \Psi^0_q}, \quad \quad \quad v = \frac{\Psi^2_1}{\sqrt{I} - \sum_q
A_{p,q} \Psi^0_q}.
\end{equation} $u$ and $v$ are parameters given by the modes before
the rotation of the reference system.  The equations for $\omega_1$
and $\omega_2$ thus obtain the form
\begin{equation} \label{B14} T_{1,0} = T_{1,1} \, u + T_{1,2} \, v,
\quad \quad \quad T_{2,0} = T_{2,1} \, u + T_{2,2} \, v.
\end{equation} These equations give the components $\omega_1$ and
$\omega_2$ and the angle $\theta$.  With the rotation matrix
(\ref{B2}) and $\omega_0=0$, we get the explicit equations
\begin{equation} \label{B15} \left\{ \begin{array}{rcl} -\sin \theta
\, \omega_2 &= & (1 - \cos \theta) \, \omega_1 \, ( \omega_1 u +
\omega_2 v ) + \cos \theta \, u,\\*[2mm] \sin \theta \, \omega_1 &= &
(1 - \cos \theta) \, \omega_2 \, ( \omega_1 u + \omega_2 v ) + \cos
\theta \, v.
\end{array} \right.
\end{equation} Multiplying the first equation with $\omega_1$ and the
second with $\omega_2$ and adding them gives the equation
\begin{equation} \label{B16} \omega_1 \, u + \omega_2 \, v = 0.
\end{equation} Together with the condition (\ref{B3}) one finds
\begin{equation} \label{B17} \omega_0=0, \quad \quad \quad \omega_1 =
\frac{v}{\sqrt{u^2+v^2}}, \quad \quad \quad \omega_2 =
-\frac{u}{\sqrt{u^2+v^2}}.
\end{equation} Inserting this into (\ref{B15}) yields the value of
$\theta$ or
\begin{equation} \label{B18} \sin \theta =
\frac{\sqrt{u^2+v^2}}{\sqrt{1+ u^2 +v^2}}, \quad \quad \quad \cos
\theta = \frac{1}{\sqrt{1+ u^2 +v^2}}.
\end{equation} With these values the rotation matrix
$T_{\alpha,\beta}$ and the transformation (\ref{B9}) and (\ref{B12})
of the modes to the new reference system are determined.


\begin{thebibliography}{2013}

\bibitem{dnapersist1} C. Bustamante, J. F. Marko, E. D. Siggia, and
S. Smith, Science {\bf 265}, 1599 (1994).

\bibitem{dnapersist2} J. F. Marko and E. D. Siggia, Macromolecules
{\bf 28}, 8759 (1995).

\bibitem{wang} M. D. Wang {\it et al.}, Biophys. J. {\bf 72}, 1335
(1997).

\bibitem{Factinpersist} A. Ott, M. Magnasco, A. Simon, A. Libchaber,
Phys. Rev. E {\bf 48}, 1642 (1993).

\bibitem{micropersist} F. Gittes, B. Mickey, J. Nettleton and
J. Howard, J. Cell Biol. {\bf 120}, 923 (1993).

\bibitem{kratky} O. Kratky and G. Porod,
Recl. Trav. Chim. Pays-Bas. {\bf 68}, 1106 (1949).

\bibitem{mods1} J. J. Hermans and R. Ullman, Physica {\bf 18}, 951
(1952).

\bibitem{mods2} H. Daniels, Proc. R. Soc. Edinburgh, Sect. A:
Math. Phys. Sci. {\bf 63}, 290 (1952).

\bibitem{mods3} N. Saito, K. Takahashi and Y. Yunoki,
J. Phys. Soc. Jpn. {\bf 22}, 219 (1967).

\bibitem{mods4} H. Yamakawa, Pure Appl. Chem. {\bf 46}, 135 (1976).

\bibitem{recent1} C. Bouchiat {\it et al.}, Biophys. J. {\bf 76}, 409
(1999).

\bibitem{recent2} J. Wilhelm and E. Frey, Phys. Rev. Lett. {\bf 77},
2581 (1996).

\bibitem{recent3} J. Samuel and S. Sinha, Phys. Rev. E {\bf 66},
050801 (2002).

\bibitem{recent4} A. Dhar and D. Chaudhuri, Phys. Rev. Lett. {\bf 89},
065502 (2002).

\bibitem{recent5} P. Gutjahr, R. Lipowsky and J. Kierfeld,
Europhys. Lett., {\bf 76}, 994 (2006).

\bibitem{recent6} B. Obermayer, O. Hallatschek, E. Frey and K. Kroy,
Eur. Phys. J. E {\bf 23}, 375 (2007).

\bibitem{liverpool1} R.E. Goldstein, S.A. Langer,
Phys. Rev. Lett. {\bf 75}, 1094 (1995).

\bibitem{liverpool2} N.-K. Lee, D. Thirumalai, Biophys. J. {\bf 86},
2641 (2004).

\bibitem{liverpool3} Y. Bohbot-Raviv, W. Z. Zhao, M. Feingold,
C. H. Wiggins, R. Granek, Phys. Rev. Lett. {\bf 92}, 098101 (2004).

\bibitem{liverpool4} T. B. Liverpool, Phys. Rev. E {\bf 72}, 021805
(2005).

\bibitem{kroy} J. T. Bullerjahn, S. Sturm, L. Wolff and K. Kroy,
Europhys. Lett. {\bf 96}, 48005 (2011).

\bibitem{winkler1} L. Harnau, R. G. Winkler and P. Reineker,
J. Chem. Phys. {\bf 104}, 6355 (1996).

\bibitem{winkler2} R. G. Winkler, J. Chem. Phys. {\bf 118}, 2919
(2003).

\bibitem{kas} J. K\"as, H. Strey and E. Sackmann, Nature {\bf 368},
226 (1994).

\bibitem{seifert} U. Seifert, W. Wintz, and P. Nelson,
Phys. Rev. Lett. {\bf 77}, 5389 (1996).

\bibitem{hall1} A. Ajdari, F. J\"ulicher, and A. Maggs,
J. Phys. (Paris){\bf 7}, 823 (1997)

\bibitem{hall2} R. Everaers, F. J\"ulicher, A. Ajdari, and
A. C. Maggs, Phys.  Rev. Lett. {\bf 82}, 3717 (1999)

\bibitem{hall3} F. Brochard-Wyart, A. Buguin, and P.-G. de Gennes,
Europhys.  Lett. {\bf 47}, 171 (1999)

\bibitem{hall4} O. Hallatschek, E. Frey, and K. Kroy,
Phys. Rev. Lett. {\bf 94}, 077804 (2005).

\bibitem{hans} G. T. Barkema and J. M. J. van Leeuwen,
J. Stat. Mech. P12019 (2012).

\bibitem{huisman} E.M. Huisman, C. Storm and G.T. Barkema,
Phys. Rev. E {\bf 82}, 061902 (2010).

\bibitem{farge} E. Farge and A. C. Maggs, Macromolecules {\bf 26},
5041 (1993).

\bibitem{expt1} C. F. Schmidt, M. B\"armann, G. Isenberg and
E. Sackmann, Macromolecules {\bf 22}, 3638 (1989).

\bibitem{expt2} A. Caspi, M. Elbaum, R. Granek, A. Lachish and
D. Zbaida, Phys. Rev. Lett. {\bf 80} 1106 (1998).

\bibitem{expt3} M. A. Dichtl and E. Sackmann, New J. Phys. {\bf 1}, 1
(1999).

\bibitem{ober2} B. Obermayer and E. Frey, Phys. Rev. E {\bf 80},
040801(R) (2009).

\bibitem{panja1} D. Panja, J. Stat. Mech. (JSTAT) L02001 (2010).

\bibitem{panja2} D. Panja, J. Stat. Mech. (JSTAT) P06011 (2010).

\end{thebibliography}
\end{document}